\documentclass[12pt,preprint]{aastex}
\usepackage{amsmath}
\usepackage{amssymb}
\usepackage{amsfonts}
\usepackage{graphicx}
\tighten
\def\lsim{\lower.5ex\hbox{$\; \buildrel < \over \sim \;$}}
\def\gsim{\lower.5ex\hbox{$\; \buildrel > \over \sim \;$}}
\def\lax {\ifmmode{_<\atop^{\sim}}\else{${_<\atop^{\sim}}$}\fi}
\def\gax {\ifmmode{_>\atop^{\sim}}\else{${_>\atop^{\sim}}$}\fi}

\def\gtorder{\mathrel{\raise.3ex\hbox{$>$}\mkern-14mu
\lower0.6ex\hbox{$\sim$}}}
\def\ltorder{\mathrel{\raise.3ex\hbox{$<$}\mkern-14mu
\lower0.6ex\hbox{$\sim$}}}
\def\pmb#1{\setbox0=\hbox{#1}\kern-0.015em\copy0\kern-\wd0
\kern0.03em\copy0\kern-\wd0
\kern-0.015em\raise0.0433em\box0 }
\slugcomment{Accepted  and scheduled for the ApJ 10 September 2004, vol. 612 issue} 
\begin{document}
\title{Spectral Index and Quasi-Periodic Oscillation Frequency Correlation in Black
Hole (BH) Sources: Observational Evidence of Two Phases and Phase Transition
in BHs }
\author{Lev Titarchuk\altaffilmark{1,2,4} and Ralph Fiorito \altaffilmark{3,4} }
\altaffiltext{1}{George Mason University/SCS/CEOSR, Fairfax VA
22030}
\altaffiltext{2}{Naval Research Laboratory, Washington
DC; ltitarchuk@ssd5.nrl.navy.mil}
\altaffiltext{3}{University of Maryland, College Park
USA; rfiorito@umd.edu; rfiorito@milkyway.gsfc.nasa.gov}
\altaffiltext{4}{NASA/ Goddard Space Flight Center, Greenbelt MD 20771;lev@lheapop.gsfc.nasa.gov}
\shorttitle{Spectral Index-Frequency Correlation} \shortauthors{Titarchuk and Fiorito}

\begin{abstract}
Recent studies have shown that strong correlations are observed between the
low frequencies (1-10 Hz) of quasiperiodic oscillations (QPOs) and the
spectral power law index of several Black Hole (BH) candidate sources, in low
hard state, steep power-law (soft) state and in transition between these
states. The observations indicate that the X-ray spectrum of such state
(phases) show the presence of a power-law component and are sometimes related
to simultaneous radio emission indicated the probable presence of a jet.
Strong QPOs ($>20\%$ rms) are present in the power density spectrum in the
spectral range where the power-law component is dominant ( i.e. 60-90\% ).
This evidence contradicts the dominant long standing interpretation of QPOs as
a  signature of the thermal accretion disk. We present the data from the
literature and our own data to illustrate the dominance of power-law index-QPO
frequency correlations.
We provide a model, that identifies and explains the origin
of the QPOs and how they are imprinted on the properties of power-law flux
component.
We argue the existence of a bounded compact coronal region which is
a natural consequence of the adjustment of Keplerian disk flow to the
innermost sub-Keplerian boundary conditions near the central object and that
ultimately leads to the formation of a transition layer (TL) between the
adjustment radius and the innermost boundary. The model predicts two phases or
states dictated by the photon upscattering produced in the TL: (1) hard state,
in which the TL is optically thin and very hot ( $kT\gtrsim$ 50 keV) producing
photon upscattering via thermal Componization; the photon spectrum index
$\Gamma\sim1.7$ for this state is dictated by gravitational energy release and
Compton cooling in an optically thin shock near the adjustment radius; \ (2) a
soft state which is optically thick and relatively cold ( $kT\lesssim$ 5 keV);
\ the index for this state, $\Gamma\sim2.8$ is determined by soft-photon
upscattering and photon trapping in converging flow into BH. \ In the TL model
for corona the QPO frequency $\nu_{high}$ is related to the gravitational
(close to Keplerian) frequency $\nu_{\mathrm{K}}$ at the outer (adjustment)
radius and $\nu_{low}$ is related to the TL's normal mode (magnetoacoustic) oscillation frequency $\nu_{MA}$. The observed correlations between index and
low and high QPO frequencies are readily explained in terms of this model. We
also suggest a new method for evaluation of the BH mass using the
index-frequency correlation.

\end{abstract}


\section{Introduction}
\label{sec:intro} The short dynamical time scales and coherence of
quasi-periodic oscillations (QPOs) observed in accreting XRB's strongly
suggest that QPOs can\textbf{\ } provide valuable information on the accretion
dynamics in the innermost parts of these systems. In this respect the
discovery of low 20-50 Hz QPOs in luminous neutron star (NS) binaries by van
der Klis et al. (1985), kilohertz QPOs in NS's by Strohmayer et al. (1996) and
hectohertz QPO's in BH's by Morgan, Remillard \& Greiner (1997) opened a new
era in the study of the dynamics near compact objects. Following these
discoveries Psaltis, Belloni \& van der Klis (1999), hereafter PBK,
demonstrated that these NS and BH low and high frequencies follow a remarkably
tight correlation. These features are the \textquotedblleft horizontal branch
oscillations\textquotedblright\ (HBO) along with \textquotedblleft low
frequency noise Lorentzian\textquotedblright, $\nu_{low}$ and the lower kHz
QPO $\nu_{high}$ for NS and BH respectively. 

 Belloni, Psaltis \& van der
Klis (2002), hereafter BPK, have updated PBK's correlation adding data from
Nowak (2000), Boirin et al. (2000), Homan et al. (2001), Di Salvo et al.
(2001) and Nowak et al. (2002). PBK suggest that the low and high frequencies
correlate in a way that seems to depend only weakly on the properties of the
sources, such the mass, magnetic field, or possibly the presence of a hard
surface in compact object. Mauche (2002) has reported low and high frequency
QPO's in the dwarf nova SS Cygni and VW Hyi (see Wouldt \& Warner 2002 for for
the VW Hyi observations) and has called attention to the fact that these QPO's
extend the correlation of PBK downward in frequency by more than two orders of
magnitude. This frequency correlation observed over such a broad range of
sources indicates that a common phenomenon is responsible for QPOs in compact
objects and raises the question  whether one can find a observational
relations which would distinguish BH candidates from other compact sources.

The comparative study of spectral and QPO features in BH candidate sources has
yielded important information which can help answer this question. As far as
spectral properties are concerned, there is increasing observational evidence
that the large complex of \ spectral \textquotedblleft
states\textquotedblright\ originally devised by Belloni (2000) to classify the
highly variable X-ray spectra of BH galactic sources like GRS 1915+105, can be
reduced to a few simple configurations [see e.g. McClintock and Remillard
(2003) and Belloni (2003)]: (1) a low hard power law dominated state where the
photon spectral index is near 1.5; (2) a steep power law state in which the
index is about 2.7; (3) and a thermally dominated state. The observed
variability is explained by transitions between these three canonical configurations.

Low frequency (1-10 Hz) QPO's are generally observed to present in states (1)
and (2), during transitions between states (1) and (2) but never in (3). Both
high and low frequency QPO's have been thought to be associated with the
accretion disk, though their specific origin has not widely understood
(although see Titarchuk \& Wood 2002, hereafter TW02, for explanation of their
origin). Attempts to find consistent correlations with observable disk
parameters have lead to mixed results. In particular, there are so many
exceptions to the observed correlations of frequency to disk spectral values
sometimes observed that no general statement about the relation of the QPO
with any spectral parameter solely associated with the disk, such as thermal
flux, temperature, inner radius, etc. can be made. For example, exceptions to
one of the most widely quoted correlations, i.e. between disk flux and
frequency, have been presented for the well studied micro-quasar GRS 1915+105
by Fiorito, Markwardt, \& Swank (2003) and\ for XTE 1550-564 \ by Remillard
et. al. (2002a,b).

On the other hand, a number of recent studies have revealed consistent and
robust correlations between the photon index of the power-law component of the
X-ray spectrum and low QPO frequency. Such strong correlations are observed
the well-studied BH binary, GRS 1915+105 in almost all situations in which the
QPO is observable and in several other BH's over a wide range of observations
and states.

These studies include:

a. A comprehensive study of correlations between photon index and QPO
frequency in BH sources by Vignarca et al. (2003); these results show strong
correlations between index and QPO frequency as well as saturation of the
index between at the values $1.7\pm0.1$ and $2.7\pm0.2$ for several BH sources
and for the same source in different states.

b. Recent studies by Kalemci (2002) for a number of galactic BH X-ray
transients observed during outburst decay which confirm the persistence of the
correlation between frequency and index during state transitions.

c. Our own studies of GRS 1915+105 of $\chi$ states (Fiorito, Markwardt, \&
Swank 2003), in which strong correlation of QPO frequency with power-law index
have been observed; such states have been traditionally noted as low
brightness, power law dominated states (see e.g. Belloni et al. 1999) where
the disk is at least tenuous if not undetectable altogether.

d. Studies tracing the QPO frequency and index over time between state
transitions shows that the frequency tracks the index during state evolution
(Homan, et. al. 2001) and Sobczak et al. (1999, 2000) from a value of about
$1.7\pm0.1$ to about $2.7\pm0.2$, as the QPO frequency varies from about 1 to
10 Hz.

The persistence of the correlation of index with QPO frequency and its tracking with respect 
to time suggests that the underlying physical process or condition which gives rise to the low frequency QPO 
are tied to the corona; and, furthermore, 
that this process varies in a well defined manner as the source progresses from one state to another. 
Moreover, the fact that the same correlations are seen in so many galactic X-ray binary BH sources, 
which vary widely in both luminosity (presumably with mass accretion rate) and state, suggests that 
the physical conditions controlling the index and the low frequency QPOs are characteristics of these sources,
 and by virtue of the low-high-frequency correlations of PBK-BPK, may be a universal property of \textit{all }
accreting compact systems.
  
The above data have motivated us to develop a detailed model of the physics of
the corona surrounding a BH s which directly predicts the behavior of the
spectral index with fundamental properties of the corona. The model we have
developed incorporates fundamental principles of fluid mechanics, radiative
transfer theory and oscillatory processes. It identifies the origin of the QPO
as a fundamental property of a compact coronal region near the BH and shows
how the photon index of this corona changes as a function of mass accretion rate.



It has been already shown in a number of papers (see below) that the absence
of the firm surface in the BHs leads to the development of the very strong
converging flow when the mass accretion rate is higher than Eddington rate
$\overset{.}{M}_{Edd}=L_{Edd}/c^{2}$. The main observational features of the
converging flow should be seen in high/soft phase, while the thermal
Comptonization spectrum of the soft (presumably disk) photons should be seen
in the hard phase.

\ In series of papers (Chakrabarti \& Titarchuk 1995, hereafter CT95; Titarchuk, Mastichiadis
\& Kylafis 1996, 1997; Ebisawa, Titarchuk \& Chakrabarti 1996; Titarchuk \&
Zannias 1998; Shrader \& Titarchuk 1998, 1999; Borozdin et al. 1999; Laurent
\& Titarchuk 1999, 2001, hereafter LT99 and LT01 respectively; Titarchuk \& Shrader 2002; Turolla, Zane \& Titarchuk
2002) the authors argue that the drain properties of black horizon are
necessarily related to the bulk inflow in BH sources and that the spectral and
timing features of this bulk inflow are really detected in the X-ray
observations of BHs. The signatures of the inflow are: 

i. an extended steep
power law where the photon index saturates to the value $2.7\pm0.2$ as the
mass accretion rate increases - the precise value of the index is a function of
the temperature of the flow but its relatively high value is a result of
inefficient photon upscattering in the converging flow due to photon
trapping. LT99 demonstrated that the spectral photon index  varies in narrow range from 3 to 2.7
with mass accretion rate (in Eddington units) increases from 2 to 7. 
Moreover, LT99 demonstrated using Monte Carlo simulations that in the low hard state when 
the plasma temperature is of order 50 keV the bulk inflow 
spectrum is practically identical to thermal Comptonization spectrum (there is no any noticeable effect 
of  the bulk Comptonization in the spectrum).  In fact, the effect of the bulk Comptonization compared
to thermal one is getting stronger when the plasma temperature drops below 10 keV.
The small variation of the photon index around  1.7 is a characteristic  signature of  the low hard state that  was pointed out  
in earlier work by CT95 and later it was confirmed by LT99. 


ii. the QPO high frequency (100-300 Hz), which is inversely proportional to BH
mass. It is worth noting that the QPO frequency scales as $1/M$ is a generic feature
of  any QPO model in which QPO frequency is scaled with the Schwarzchild radius.


Furthermore, Titarchuk, Osherovich \& Kuznetsov (1999), hereafter TOK,
presented observational evidence that the variability of the hard X-ray
radiation in NS's and BH's occurs \textit{in a bounded configuration}. Ford \&
van der Klis (1998) found that the break frequency in the power density
spectrum (PDS) is correlated with QPO frequency for some particular NS sources
(4U 1728-32). Wijnands \& van der Klis (1999) later found a similar
correlation in BH sources. Titarchuk \& Osherovich (1999), hereafter TO99, and
TOK explained this correlation in NSs and BHs respectively. They argue the
observed PDS is the power spectrum of an exponential shot, $\exp(-t/t_{d})$
which is the response of the diffusion propagation to any perturbation in a
bounded medium. The inverse of the characteristic diffusion time $\nu
_{b}=1/t_{d}$ and the normal mode QPO frequency $\nu_{0}$ has a well defined
relation to the size of the bounded configuration, because $\nu_{b}\propto V/L$ and 
$\nu_{0}\propto V/L^{2}$ where $V$ is the specific pertrubation
(magneto-sonic) velocity and $L$ is a charactestic size of the configuration
(see TO99).

Titarchuk, Lapidus \& Muslimov (1998), hereafter TLM98 proposed that this
bounded configuration (cavity) surrounding compact objects is the transition
layer (TL) that is formed as a result of dynamical adjustments of a Keplerian
disk to the innermost sub-Keplerian boundary conditions. They argued that this
type of adjustment is a generic feature of the Keplerian flow in the presence
of the sub-Keplerian boundary conditions near the central object and that it
does not necessarily require the presence or absence of a hard surface. TLM98
concluded that an isothermal sub-Keplerian transition layer between the NS
surface and its last Keplerian orbit forms as a result of this adjustment. The
TL model is general and is applicable to both NS and black hole systems. The
primary problem in both NS and BH systems is understanding how the flow
changes from pure Keplerian to the sub-Keplerian as the radius decreases to
small values. TLM98 suggested that the discontinuities and abrupt transitions
in their solution result from derivatives of quantities such as angular
velocities (weak shocks). They were first to put forth the possibility of the
transition layer formation to explain most observed QPOs in bright low mass
X-ray binaries (LMXBs). In Figure 1 we illustrate the main idea of the
transition layer concept for NS's ans BH's.


In this \textit{Paper}, we explain the general correlation between photon index and low
frequency   in the frameworks of the transition layer model and
we give more arguments for the nature of the spectral phases (states) and
phase transition observed in BHs. The main features of the TL model are given
in \S 2.1 The formulation of the problem of low frequency oscillations and the
relationship with MA oscillations in the transition layer are also  described in
\S 2.1. The coronal model and the spectral index-optical depth relation is given in \S 2.2. 
The derivation of the low frequency-index correlation 
and the details of modeling of the spectral phase  transition are  present in
\S 2.3.
  We  analyze specific QPO and spectral data in
terms of the TL model in \S 3. We discuss the signatures and the methods of the
identification of BH and NS sources using timing and spectral characteristics
in \S 4. Our summary and conclusions also follow in \S 4.

\section{Transition Layer Theory}

\subsection{The main features of the transition layer model}

TLM98 define the transition layer as a region confined between the the inner
sub-Keplerian disk boundary and the first Keplerian orbit (for the TL
geometry, see Fig. 1 and Fig. 1 in TLM98 and TOK). 
The main idea at the basis of our investigation is that a Keplerian
accretion disc is forced to attain sub-Keplerian rotation close to
the central object (a neutron star or a black hole). The
transition between the Keplerian and the sub-Keplerian flow takes
place in a relatively narrow, shock-like region [the size of the region of order 
radius of the central object (NS, BH)]  where dissipation occurs. As
a consequence, the gas temperature increases and the disc puffs
up, forming a hot corona. Then the corona  intercepts the soft
disc photons, up-scatter them via thermal and dynamical
Comptonization finally giving rise to the hard power-law tail. 

The  power-law index value 
strongly depends on the coronal  optical depth and  temperature (see  CT95, TMK97,  LT99).
When the mass accretion rate in the disk increases  and consequently the disk soft photon flux increases, 
the corona is drastically cooled down to a temperature of order 5-10 keV 
(in \S 2.2 we present a detailed analysis of this effect). For such a low plasma temperature 
the bulk motion of the converging flow is more efficient in upscattering disk photons than thermal Comptonization.
Furthermore, the indices saturates to the asymptotic values around 2.7 as the mass accretion rate of the converging 
flow increases. The exact asymptotic value of the index is determined by the plasma temperature only (LT99).    
 
TLM98 evaluate the size of the transition layer $L$ as a function of
a nondimensional paramerer $\gamma$ (often called the Reynolds number) which is the inverse of
the  $\alpha-$ parameter for the accretion flow (Shakura \& Sunyaev 1973, hereafter SS73),
\begin{equation}
\gamma=\dot{M}/4\pi\eta H=RV_{r}/D, 
\label{eq:rey}
\end{equation}
where $\dot{M}=4\pi RH m_pN_eV_{r}$ is the accretion rate in the disk,  $H$ is a half-thickness of a disk,   
$V_{r}$ is a characteristic radial  velocity  at a given radius $R$ in the disk,  $N_e$  is electron (proton) number density , $m_p$ is  the proton mass 
 and $D$ is the diffusion
coefficient. $D$ can be defined as $D=V_{t}l_{t}/3$ using the turbulent velocity
and the related turbulent scale, respectively or as $D=D_{M}=c^{2}/\sigma$ for
the magnetic case where $\sigma$ is the conductivity (e.g. see details of the
$D-$definition in Lang 1998). It is worth noting that  the viscosity $\eta$ is a function of $\dot M$.  

Because $\gamma$ is related to  $\dot{M}$ and the TL  Thomson optical depth 
$\tau_0$,  which is  a product  of  $L=R_{out}-R_{in}$ ($R_{in}$, $R_{out}$ are the TL inner and the outer radius 
respectively), electron number density $N_e$ and Thomson cross-section $\sigma_{\rm T}$ 
(namely, $\tau_0=\sigma_{\rm T}N_eL$)  are both related to  $\dot{M}$ one can,  in principle,  determine $\tau_0$  
as a  function of $\gamma-$parameter.  
Because of the uncertainties of the disk equation of the state, of the viscosity $\eta$ in the disk 
there is still uncertainty in the precise determination of this $\tau_0(\gamma)-$relation. 
We address to this issue 
in \S 2.3.

TLM98 determine the $L(\gamma)-$dependence using the equation of motion in the disk 
where the radial motion is controlled by friction 
and the angular momentum exchange between adjacent layers, resulting 
in the loss of initial angular momentum by the accreting matter (e.g. SS73). 
We define an adjustment point where a Keplerian accretion disc is forced to attain 
sub-Keplerian rotation close to the central object (NS or BH). 
In order to determine the adjustment radius one should solve this equation of 
motion subject to three boundary conditions~--~one at the innermost disk radius $R_{in}$ and 
two at the adjustment radius $R_{out}$~--~namely, at $R_{in}$ the disk rotational frequency matches
 the rotational frequency of the central object (spin), at $R_{out}$ there is no break in the rotational frequency 
 and the left-hand and right-hand side derivatives of  R are equal.
Thus, for a given value of $\gamma-$parameter the rotational frequency profile $\omega(R)$ and the
outer radius of the transition layer are uniquely determined by the three boundary
conditions and the  equation of motion (see for example, Eqs. 8-9 in T099).

The adjustment of the Keplerian disk to the sub-Keplerian inner boundary creates
conditions favorable for the formation of a hot plasma outflow at the outer
boundary of the transition layer, because the Keplerian motion (if it
is followed by sub-Keplerian motion) must pass through the super-Keplerian
centrifugal barrier region (TLM98).    
 

The equation (see TLM98, Eq. 11, or TO99, Eq. 9)
determines $R_{out}$
as a function of $\gamma$. 
For given values of $R_{out}(\gamma)$  one can find values of the Keplerian frequencies
$\nu_{\rm K}(R_{out})$
(see Eq. 1 in TO99).  
For example for the observed range of the high QPO frequencies  in BH sources, 
 from 1 to 100 Hz (the low QPO frequencies from 0.1 to 10 Hz)
(see Fig. 2), and  for a BH mass of ten solar masses one can find from TLM98 Eq. (11)
that $\gamma$ varies from 2 to 20 (a variation which is related to the mass accretion rate $\dot{M} $
see Eq. \ref{eq:rey}).  We remind a reader that in the framework of the TLM  
these high QPO frequencies are interpreted as Keplerian frequencies $\nu_{\rm K}$ at $R_{out}$.

It is worth noting that in the literature $\alpha=\gamma^{-1}$ is treated
as a constant and a free parameter independent of the mass accretion rate (or
the spectral state), whereas we find that  $\alpha$ varies from 0.5 to 0.05
with $\dot{M}$, i.e. $\alpha$ is a very strong function of $\dot{M}$. 

Soria (1999) inferred the effective value of the viscosity $\alpha_{eff}-$parameter as a function of the magnetic field.
He showed that regardless of the true viscosity in the disk the  $\alpha_{eff}-$ parameter
 is always higher than $0.05$ if the ratio of the magnetic pressure $P_m$
to the gas pressure $P_g$  is about 10\% and more.    
Below (see also TW02 for details) we show  that using the low frequency-high frequency correlation
the inferred ratio of $P_m/P_g=1/\beta=0.1$ that is consistent with  the values of 
$\alpha=0.05-0.5$ and the Soria's  predictions of the high values of $\alpha$ as
$P_m/P_g\gax0.1$


The TL model identifies the low frequency QPO as that associated with the
viscous magnetoacoustic (MA) oscillation of the bounded TL; this mode is
common to both BH's and NS's.
The correlation of the MA frequency with the gravitational (Keplerian)
frequency $\nu_{\mathrm{K}}=2\pi\omega_{\mathrm{K}}$ were derived by TO99,
Titarchuk, Bradshaw \& Wood (2001) for NS and then it was generalized by 
TW02,
for BH and WD. The MA frequency is derived as the eigenfrequency of the
boundary-value problem resulting from a MHD treatment of the interaction of
the disk with the magnetic field. The problem is solved for two limiting
boundary conditions which encompass realistic possibilities. The solution
yields a velocity identified as a mixture of the sound speed and the Alfv\'en
velocity.
The TBW treatment does not specify how the eigenfrequency is excited or
damped. However, it makes clear that the QPO is a readily stimulated resonant
frequency [see Titarchuk, Cui \& Wood (2002) and Titarchuk (2002) for details
of excitation of the eigenfrequencies].


A linear relation was derived between $\nu_{MA}$ and $\nu_{\mathrm{K}}$ (see
Fig. 2 and TW02)
\begin{equation}
\nu_{MA}=C_{MA}\nu_{\mathrm{K}}.
\label{eq:MA}
\end{equation}
where $C_{MA}=2^{1/2}4\pi\lbrack(1+f\beta)/(1+\beta)]^{1/2}(H/r_{out})$ is a
proportionality coefficient which is \textquotedblleft
universal\textquotedblright\ to the extent that $\beta$ and $H/R_{out}$ remain
about the same from one source to the next. Here $\beta=P_{g}/P_{m}$ is the
ratio of the gas pressure to the magnetic pressure and the coefficient
$f=1/32,~1/32\pi^{2}$ for stiff and free boundary conditions respectively and
$H$ is the half-width of the Keplerian part of the disk.

Identification of $\nu_{high},~\nu_{low}$, with $\nu_{\mathrm{K}}, ~\nu_{MA}$
respectively, leads TW02 to determination of $H/R_{out}=1.5\times10^{-2}$ and
$\beta=10$ using the observed value of $C_{MA}=0.08$.
On the other hand $\nu_{\mathrm{K}}=\nu_{high}$ at the adjustment radius $R_{out}$ as of function of 
$\gamma-$parameter can be readily obtained  using Eq. (9) in TO99 [see above 
for details of $\nu_{\mathrm{K}}(\gamma)-$ determination].
Thus, as a result of  the identification of $\nu_{low}$ with $\nu_{MA}$, we obtain the  relation 
of $\nu_{low}=\nu_{MA}=C_{MA}\nu_{\rm K}$ as a function of $\gamma$, namely
\begin{equation}
\nu_{low}=
0.08\nu_{\mathrm{K}}(\gamma).
\label{high-low}
\end{equation}
In Figure 3 we present the inferred $\nu_{low}$ 
as a function of $\gamma$. 
We derive this relation  using equations (9) in TO99 
with the assumptions that $a_{\mathrm{K}}=m(\nu_{0}/363~Hz)=1$ 
(see Eq. 7 in TO99 for definition $a_{\rm K}$) and $m=M/M_{\odot}=10$.

It is evident that the inferred $\nu_{high},~\nu_{low}$ are inversely
proportional to $m$ because
\begin{equation}
\nu_{high}=\nu_{\mathrm{K}}\propto m^{-1}%
\end{equation}
and
\begin{equation}
\nu_{low}\sim V/L,
\end{equation}
where $L$ is the TL characteristic size which is proportional to $m$.

We calculate the outer radius of the TL, $r_{out}=R_{out}/R_{\mathrm{S}}$ as a
function of $\nu_{low}$ using relations [Eq.\ref{eq:MA} (or Eq. \ref{high-low}) here and Eqs. 1, 9 in TO99], 
keeping in mind that $\nu_{high}=\nu_{\mathrm{K}}$. 
Because the radius of interest, which in our case is the outer TL boundary
radius $R_{out}$, is related to $\nu_{\mathrm{K}}$ one can relate $R_{out}$ to
$\nu_{low}$ using formula (\ref{high-low}). We present this function in Figure 4.

The spectral index$-$frequency correlation can be derived if we find a
spectral index$-$TL optical depth correlation. Because $\gamma$ is
is a function of the mass accretion rate $\dot{M}$ and the optical depth
$\tau_{0}$ is a function of $\dot{M}$ , \ we are lead to a correlation
of $\nu_{low}$ and $\tau_{0}$ and ultimately to the index$-$frequency correlation.
It is worth noting that $\gamma$ and $\tau_0$ dependences on $\dot M$ implies that  a correlation between these
quantities exist  {\it but does not mean a linear correlation between $\gamma$ and $\tau_0$ because $\gamma-$parameter
also depends on the accretion flow viscosity $\eta$} which  is also a function of $\dot M$ .

\subsection{Spectral index-optical depth relation. Corona model}

The TL model first proposed by TLM98 naturally provides a compact corona for Comptonization upscattering of 
the illuminating accretion disk photons which is just the region of adjustment of Keplerian accretion flow (disk)
 to the sub-Keplerian innner boundary condition. 
The essential sub-Keplerian rotation of NS in LMXBs is a well established observational fact due 
to the direct measurements of NS spins 
(which are in the range of 200-400 Hz, see the review by van der Klis 2000 for details). 
Thus it is quite reasonable that the presence of a TL for NSs exists. 
In the case of BHs no direct spin measurements are yet available and hence the presence of the TL may 
at first glance seem less clear.  
On the other hand, the Bulk Motion Comptonization model which requires the existence of the innermost region 
of non Keplerian flow onto a slowly rotating or static Schwarschild BH, is very consistent with observations and 
in fact its signature is the saturation of the photon index which is observed (see LT01 for details) and thus 
one can conclude that fast rotating Kerr BH is  ruled out by the observations.  
In fact, there is no  space between the the inner edge of the disk and the horizon 
in the fast rotating Kerr  BH to upscatter the disk soft photons by the bulk motion
to the energies of order 500 keV that are detected in the high energy observations (Grove et al. 1998).
 Thus we can reasonably conclude that BH may also undergo sub-Keplerian rotation and the existence of a TL for slowly
 rotating BHs. 

It is
likely that this adjustment to sub-Keplerian flow is not smooth and that near
the adjustment radius the strong or weak shocks can be formed.
The adjustment radius (shock)
region can be treated as a potential wall at which the accreting matter
releases its gravitational energy in a geometrically thin target. As a
consequence of this the shock plasma temperature is much higher than that in
the surrounding regions of the Keplerian disk. The shock region gets puffed up
oscillating with frequencies close to $\nu_{\mathrm{K}}$ (Titarchuk 2003).
Additional oscillations can appear (at least in NS case) because of the
rotational configuration above the disk which is a magnetosphere in NS case
(see details in TOK). The shock formation leads to the formation of the hot
inner bounded region. The hard photons in the shocked region illuminate the
surrounding material evaporating some part of the disk. It is obvious that the
evaporated fraction depends on the mass accretion rate in the disk (see CT95). Some small fraction of the
hard photons (at maximum 6 \%) can be reflected by the cold parts of the disk
(CT95 and see also Basko, Sunyaev \& Titarchuk 1974).

Let us assume the column density of this region $y_{0}$ is a few grams or that
the Thomson optical thickness $\tau_{0}$ is a few (see SS73 for the thickness
estimate of the innermost part of the disk). The rate of energy release at
this region $Q_{cor}$ is a few percent of \ the Eddington luminosity since the
adjustment radius is located within 4-100 Schwarzschild radii (see Fig. 4).
The heating of gas due to the gravitational energy release should be balanced
by the photon emission. For high plasma temperature, Comptonization and
free-free emission is the main cooling channel, and the heating of electrons
is presumably due to the Coulomb collisions with protons (see Fig. 1 for the
picture of disk corona).

Under such physical conditions the energy balance can be written as [see also
Zeldovich \& Shakura (1969), hereafter ZS69 and  TLM98]
\begin{equation}
Q_{cor}/\tau_{0}=C_{comp}\varepsilon(\tau)T_{e}/f(T_{e})+C_{ff}T_{e}^{1/2}%
\rho,
\label{eq:ZS}
\end{equation}
where $\tau$ is the optical depth within the shock,
$\varepsilon(\tau)$ is a distribution function for the radiative energy
distribution, $f(T_{e})=1+2.5(kT_{e}/mc^{2})$ is the relativistic correction
factor, $T_{e}$ is the electron temperature in K, $C_{comp}=20.2$ cm s$^{-1}$
K$^{-1}$ and $C_{ff}=2.6\times10^{20}$ erg K$^{-1/2}$cm s$^{-1}$ are a
dimensional constants.
In this formula we neglect the gas heating due to recoil effect.

The distribution $\varepsilon(\tau)$ is obtained from the solution of the
diffusion equation (ZS69),
\begin{equation}
d^{2}\varepsilon/d\tau^{2}=-3(Q_{tot}/c)/\tau_{0}
\label{eq:diff}
\end{equation}
subject to two boundary conditions. Here we assume that the region of the
gravitational energy release (corona) is a spherical shell surrounding the
central object and the total flux in the corona is a sum $Q_{tot}
=Q_{cor}+Q_{d}$ of the gravitational energy release $Q_{cor}$ and the
illumination flux from outside of the corona (disk) $Q_{d}$. The inner and
outer boundary conditions for a BH corona are that there are no scattered
radiation from outside of the corona. Then in the Eddington approximation (see
Sobolev 1975) the boundary conditions can be written as follows :
\begin{equation}
(d\varepsilon/d\tau-3\varepsilon/2)|_{\tau=0}=(d\varepsilon/d\tau
+3\varepsilon/2)|_{\tau=\tau_{0}}=0,
\label{eq:bdcon}
\end{equation}
where $\tau=0$ and $\tau=\tau_{0}$ are at the inner and outer boundaries respectively.

For a NS corona one should assume that inside of the inner corona boundary the
energy density $\varepsilon=$constant, namely that the radiation flux emitted
towards the central object at some point of the inner boundary returns back
(reflected by the NS surface), namely this condition is equivalent to the
reflection condition:
\begin{equation}
d\varepsilon/d\tau|_{\tau=0}=0.
\end{equation}
Below we present a general formula of the coronal temperature which combines
the BH and NS cases.

The solution of equations (\ref{eq:diff}-\ref{eq:bdcon}) provides us with the distribution function
for the energy density in the BH corona
\begin{equation}
\varepsilon(\tau)=(Q_{tot}/c)\{1+(3/2)\tau_{0}[\tau/\tau_{0}-(\tau/\tau
_{0})^{2}]\}.
\end{equation}
We neglect the dependence of $Q_{tot}$ on $\tau$ to derive this formula.
Furthermore, we take a representative value of $<\varepsilon>$ for the maximum
value in the corona which is very close to mean value of $\varepsilon_{mean}$
i.e.
\begin{equation}
\varepsilon>=\varepsilon_{max}=(Q_{tot}/c)[1+(3/8)\tau_{0}].
\end{equation}
in order to approximate the coronal temperature. In Figure 5 we present the
results of calculations of the temperatures $kT_{e}$ (keV) and the energy
spectral indices of the Comptonization spectrum $\alpha$ (photon index
$\Gamma=\alpha+1$) as a function of the optical depth of the shell $\tau_{0}$.
The different curves of $T_{e}(\tau_{0})$ are related to the different ratios
of $Q_{d}/Q_{cor}$. The calculations have been made for values of
$\rho=10^{-6}$ g cm$^{-3}$ and $Q_{cor}=10^{22}$ erg cm$^{-2}$ s$^{-1}$ which
are characteristic values of the density and luminosity for the standard disk
model (SS73).
In fact, the calculation results weakly depend on $\rho$ and $Q_{cor}$ if
\[
(\rho/10^{-6}\mathrm{{g~cm}^{-3})/(Q_{cor}/10^{22}erg~cm^{-2}s^{-1})\lax 1 }%
\]
is
of order a few or less.
In this case the equation for the temperature is simplified
\begin{equation}
6.73\times10^{-2}\tau_{0}(1+3\tau_{0}/8)(T_{e}/10^{8}~K)(1+Q_{d}
/Q_{cor})/f(T_{e})=1
\label{eq:temp}
\end{equation}
which has the solution
\begin{equation}
T_{e}/10^{8}~K=1.5\times\lbrack\tau_{0}(1+3\tau_{0}/8)(1+Q_{d}/Q_{cor}
)-0.62]^{-1}.
\label{form:temp}
\end{equation}
It is easy to show that the case for which $Q_{d}/Q_{cor}=1$ is identical to
the NS case (the inner reflective boundary) when the disk illumination is
neglected. One can see from Eq. (\ref{form:temp}) that $T_{e}$ is a function of $\tau_{0} $ (only)
 if $Q_{d}/Q_{cor}\ll1$. It is also worth noting that $T_{e}\tau_{0}(1+3\tau_{0}/8)/f(T_{e})$ 
 is insensitive to the total luminosity $Q_{tot}$ if $Q_{d}/Q_{cor}\ll1$ (Eq. \ref{eq:temp}). 

The spectral index $\alpha$ as a function of the product of $T_{e}\tau_{0}$
are insensitive to $Q_{tot}$ too for $\tau_0\lax1$ because the the
Comptonization parameter
\[
y=4kT_{e}/m_{e}c^{2}~\mathrm{Max}(\tau_{0},\tau_{0}^{2})
\]
and
\[
\alpha=-3/2+\sqrt{(9/4+4/y)}
\]
(see Rybicki \& Lightman 1979 and Sunyaev \& Titarchuk 1980). This inferred
property of the Comptonization spectra reproduces the observed independence of
$\alpha$ on the bolometric luminosity (see e.g.  Tanaka
1995) when $\alpha$ varies within the range $0.6\pm0.2$ ($\Gamma=1.6\pm0.2$).
This defines the so called the low/hard state in BHs.

Below we show that the observed plateau of the index-frequency correlation at
low values of $\nu_{low}<1$ Hz is the result of this behaviour because the lower
frequency values are related to the low mass accretion rates $\dot{m}=\dot{M}/M_{Edd}<1$ and 
$\tau_{0}$ is of order one when $\dot{m}<1$.

We calculate the exact values of the spectral indices $\alpha$ for a given
$T_{e}$ and $\tau_{0}$ using the relativistic formulas developed by Titarchuk
\& Lybarskij (1995) [see formulas (17), (for plane geometry, note their
$\tau_{0}$ is our $\tau_{0}/2$) and (24) there]. In Figure 5 we present the
spectral indices as a function of $\tau_{0}$ for BH and NS cases when the disk
illumination is negligible small with respect to the coronal energy release (a
condition which can be a model for the low/hard state in these systems). The
plasma temperature values of 20-150 keV for $\tau_{0}=1-3$ and $\alpha
=0.6\pm0.2$ (for $ratio=Q_{d}/Q_{cor}=0$) are typical values of these
quantities for low/hard state in BHs.

When $Q_{d}$ is comparable with $Q_{cor}$ the cooling becomes more efficient
due to Comptonization and free-free processes and therefore $T_{e}$
unavoidably decreases [see $T_{e}-$values for $Q_{d}/Q_{cor}=(1-4)$ in Fig. 5].

This illumination effect and mass accretion rate increase may explain the
hard-soft transition when the Compton temperature drops substantially with
increase of the soft (disk) photon flux (for illustration, see Figure 1 ). In
Figure 5 we show that temperature drops from 50-60 keV for $ratio=0$ to 7-10
keV for $ratio=4$. The temperature may drop also due to an increase in the
optical depth (presumably because $\dot{m}$ increases) even when $ratio=0$.

\subsection{Modeling of spectral phase transition from the hard state to the
soft state}

LT99  studied the Comptonization of
the soft radiation in the converging inflow (CI) into a black hole using Monte
Carlo simulations. The full relativistic treatment has been implemented to
reproduce the spectra. 
LT99 show that spectrum of the soft state of BHs can be described as the sum
of a thermal (disk) component and the convolution of some fraction of this
component with the CI upscattering spread (Green's) function. The latter
boosted photon component is seen as an extended power law at energies much
higher than the characteristic energy of the soft photons. LT99 demonstrate
the stability of the power law index (the photon index, $\Gamma=2.8\pm0.1$)
over a wide range of the plasma temperature 0-10 keV, and mass accretion rates
(higher than 2 in Eddington units) due to upscattering and photon trapping in
the converging inflow. The spectrum is practically the same as that  produced
by standard thermal Comptonization
 when the CI plasma temperature is of order 50 keV (the typical ones for
the BH hard state) and the photon index  $\Gamma$ is around 1.7

LT99 also demonstrate that the change of the spectral shapes from the soft
state to the hard state is clearly related to the temperature and optical
depth of the bulk inflow. We combine our results of calculations of index and
Compton cloud temperatures for thermal Comptonization case (see \S 2.2 and Fig.5) and 
LT99's results of the spectral calculations (see Table 2 in LT99 for
the spectral index values) to evaluate the power law index as a function of
the optical depth of the Compton cloud $\tau_{0}$. We use the values of the
index and of the temperature for $Q=0$ (see Fig. 5) when $\tau_{0}$ varies
from 1 to 3 to decribe the index behavior and the plasma temperature variation
in the hard state. Figure 6 presents the photon index $\Gamma$ as a function
of $\tau_{0}$. This function $\Gamma(\tau_{0})$ can be fitted analytically as
follows:
\[
\Gamma(\tau_{0})=c_{0}+c_{1}\tau_{0}+c_{2}\tau_{0}^{2}+c_{3}\tau_{0}%
^{3}~~~~~\mathrm{for}~~~~1<\tau_{0}~\leq~5.17,
\]%
\begin{equation}
\Gamma(\tau_{0})=b_{0}-b_{1}e^{-\tau_{0}/\tau_{\star}}~~~~~\mathrm{for}%
~~~~\tau_{0}~>~5.17
\label{eq:gam-tau}
\end{equation}
where $c_{0}=0.61,~c_{1}=0.57,~c_{2}=1.485\times10^{-2},~c_{3}=-8.5\times
10^{-3}$ and $b_{0}=2.81$, $b_{1}=1.0\times10^{3}$ and $\tau_{\star}=0.5$.

In order to derive the index dependence $\Gamma$ on $\nu_{low}$ one should
relate the parameter $\gamma$ to $\tau_{0}$ and then combine the two
derived relationships $\Gamma$ vs $\tau_{0}$ and $\nu_{low}$ vs $\gamma$ (see Figures
3-6 ). The optical depth $\tau_{0}$ and $\gamma-$parameter are functions of  $\dot{M}$. 
To fit the data we
assume that $\tau_{0}=A\gamma^{\delta}$. We have three free parameters, black
mass $m=M/M_{\odot}$, $\delta$ and $A$ to fit the model to the data if we
assume that the converging inflow (CI) is essentially cold (the CI temperature
$kT_{ci}\lax5$ keV). This relationship is all that is needed to fit the observed $\Gamma-\nu_{low}$
correlations.


\section{Data and Analysis}

We analyze data points for photon index vs low frequency correlation obtained
from Trudolyubov et al. (1999) and Vignarca et al. (2003) for GRS 1915+105 and
Sobczak et al. (1999, 2000) for XTE J1550-564. Also we make use 
data for GRS 1915+105 by Fiorito, Markwardt \& Swank (2003) hereafter FMS,
which is a detailed study of QPO and spectral features of the steady low hard
states of GRS 1915+105 for representative 1996 and 1997 observations. FMS
computed the power spectra of combined Binned and Event Mode data for PCA
channels 0-255 and sampled at 0.25-64 Hz for 4-second intervals. The spectra
were averaged over the entire observation duration. The PDS spectra were fit
with a combination of Lorentzian lines and exhibited a break frequency, a
central QPO frequency and in some case the presence of the first harmonic.
Background subtracted energy of each OBSID were obtained and time average over
the entire observation duration using Standard 2 PCA and HEXTE data where
available (both clusters). The XSPEC model used to fit the energy spectra
consisted of a multi-temperature thermal disk component (DISKBB), a simple
power law and a small Iron line component. For the 1996 data (OBSID10408), it
was found that the disk component could be neglected without affecting the
fit. Acceptable $\chi^{2}$ were obtained in all cases.

\subsection{Correlations between QPO Frequency and Spectral Parameters}

The first part of the FMS study was to first to look for correlations of
different spectral components, presumably thermal disk components, with QPO
frequency to try to determine the reason for the non-unique total flux
correlations previously observed. In the case of OBSID 20402 (1997) we did
observe the expected positive correlation of QPO frequency with disk flux in
agreement with previous analyses of Muno et al. (1999) and Markwardt et al.
(1999), which were done in other states. However, we observed an
anti-correlation of frequency with disk flux for OBSID 10408 (1996); and no
correlation of frequency with disk flux for OBSID 10258 (1996). Furthermore,
no correlation of QPO frequency with disk temperature is seen for any of the
OBSIDS we studied. These results are at variance with the often-quoted result
that disk parameters are closely associated with QPO frequency.
Remillard, et. al. (2002b) show a comparison of low QPO frequency and the
apparent disk flux for the source XTE J1550-564 during its 1998-1999 outburst.
The QPOs are reported with differences in phase lags. Again, it is clear that
a frequency-disk flux correlation is observed for this source is only observed
at lower ($\nu<5$ Hz) QPO frequencies and fluxes than usually observed for GRS 1915+105.



However, robust correlation between index and the central frequency of low
frequency QPOs are also seen in a wide variety of BH sources radiating in
different states and during transition between states. The QPO frequencies in
such studies are observed in the PDS obtained by integrating over time
intervals ranging from the complete time of the observation to segmented time
bins where the luminosity remains constant, [see e.g. Markwardt, et. al
(1999)]. Strong frequency-index correlations are observed by Kalemci (2002) in
outburst decay (presumably state transitions) of a number of sources and
by Vignarca, et. al. (2003), in a comprehensive study which focuses its
attention on the behavior of the power law index and QPO frequency in several
BH candidate sources. We will use the data of Vignarca, our studies and that
of other as well in our analysis (see \S 3.2 below).

Observation of low term tracking of the QPO frequency with the variations of
the photon index has been observed in at least two BH sources. Tracking is
seen in the data of Rossi (2003) and Homan (2001) in observations of XTE
J1560-500 taken over a 30-day period. In this period the source apparently
transits from the low hard state ($\Gamma\sim1.7$) to the intermediate state and  the soft
state where the index apparently remain close to the constant at a value (
$\Gamma\sim2.7$ ). In this transition the QPO frequency varies from about 1 to
10 Hz.
Similar tracking is observed in XTE J1550-564 by Sobzcak et al. (1999), (2000). However, the
index and QPO frequency does not track with the total X-ray flux (Vignarca et
al. 2003, Figure 7).

\subsection{Comparison of the inferred index-frequency correlation with the
data. Low/Hard-High/Soft State Transitions}




\centerline{\it GRS1915+105}
In Figure 7 we present the results of our fits to the data of Vignarca, et. al (2003) 
for the plateau observations of the microquasar GRS 1915+105 using the TL model.
The best fit parameters are $m=12$, $\delta=1.25$ and $A=1$. We note
that the value $m=12$ is consistent with the BH mass evaluation $m=13.3\pm4$
obtained by Greiner et al. (2001) using the IR spectra and by Shrader and
Titarchuk (2003) (see also Borozdin et al. 1999) using the X-ray spectra. The
observable index saturations at low and high values of $\nu_{L}$ are nicely
reproduced by the model. The low and high frequency plateau regions of the
data and fit are signatures of the two spectral phases which are explained by
two different regimes of Comptonization ( upscattering): \ (1) bulk flow
Comptonization in the soft state ( saturation at $\Gamma\sim 2.8$) and
(2) thermal Comptonization in the hard state ( the photon index tends to level  at $\Gamma
\sim1.7$).

A comparison of the index-frequency correlation of plateau observations with
Vignarca's $\beta$ and $\nu$ data [Belloni (1999) classification scheme] and
the model fits are plotted in Figure 8. The only difference in the data is
different saturation values at high $\nu_{L}$ that can be readily explained by
a change of plasma temperature in the converging inflow $kT_{ci}$. For upper value of
$\Gamma=2.81$ the temperature $kT_{ci}=5$ keV as for $\Gamma=2.7$ the
temperature $kT_{ci}=9$ keV. The best-fit parameters for the latter data are
the same as for the plateau observations except that in the model function
$\Gamma(\tau_{0})$ (see Eq. \ref{eq:gam-tau}) the coefficents $b_{0}$, $b_{1}$ and
$\tau_{\star}$ are replaced by $b_{0}=2.7$, $b_{1}=22.4$ and $\tau_{\star}=0.8$ for $\tau_{0}>2.47$.

In Figure 9 we present a plot of power-law index versus QPO centroid frequency
for the observations of class $\alpha$ and $\nu$ of GRS 1915+105 from Vignarca
et al. (2003) (black points=obs. 18). Blue points correspond to the values for
observations by Fiorito et al (2003) for OBSID 20402 and OBSID 10258. Magenta points
correspond to positions where the OBSID 20402 data points are shifted to one half the measured QPO
low frequency values. This suggests that the latter data points are related to
the second harmonics of the $\nu_{L}$ frequencies. It is well known that data
for NS QPOs is a mixture of the first and the second harmonics of horizontal
branch oscillation (HBO) frequencies (see e.g. van der Klis 2000). Red points
correspond to positions where should be points with half of frequencies for
Vignarca' s obs. 18. 
A theoretical curve (blue solid line)
obtained using $\Gamma(\tau_{0})$ (see Eq. \ref{eq:gam-tau}) (for coefficents $b_{0}=2.7$,
$b_{1}=22.4$ and $\tau_{\star}=0.8$ for $\tau_{0}>2.47$) for $m=12$ and
$\tau_{0}=\gamma^{1.5}$ fits the FMS OBSID 10258 data points.

It is  worth noting that relation between $\gamma$ and $\tau_0$ is slightly different from that
we obtained for the best-fits presented in Fig. 8.
 Namely, the index of $\gamma$, increases from 1.25 to 1.5.
It means that for  the same value of  $\gamma$ the  optical depth
 $\tau_0$ is slightly higher than that in  state presented in Fig. 8.
This change can be explained by a  slightly higher accummulation of plasma  in the transition layer. 
For example, if  we  fix the disk viscosity  $\eta$, then   the $\gamma-$parameter is determined by
$\dot M$ only. But the mass accretion rate $\dot M$ depends on the product of number density $N_e$ 
and the flow radial velocity $V_r$, 
i.e $\dot M\propto N_eV_r$. Thus, for a given value of $NV_r$ the optical depth $\tau_0\propto N_e$ varies as
 $V_r$ varies. Smaller values of $V_r$ corresponds to the higher values of $N_e$ and consequently to higher values of $\tau_0$, 
 namely  higher plasma accummulation in the transition layer can be related to smaller velocities $V_r$. 
 

It is interesting to note that  the suggested half-frequency points (magenta and red ones) are 
located in a narrow corridor between these  inferred curves 
(presented  in Fig. 8 and Fig. 9 respectively). 

\centerline{\it XTE J1550-564}
 
Figure 10 shows a plot of power law index versus QPO frequency for XTE
J1550-564.  All data points presented in  Figures 10, 11 see also 
 (see also  Fig. 6, 8 in Vignarca et al.) were obtained by
 Sobczak et al. (1999), (2000); Remillard et al. (2002a,b).  

The data again show a flattening out or saturation of the index at a value $\Gamma\sim2.8$. It is clearly seen from
Fig. 8 of Vignarca et al. that most of the points presented there can be
fit by a smooth line except for  a few points on the right that we temporarily exclude from our
consideration. The shift transformation $\nu_{low}^{\prime}
=12\nu_{low}/10$ of the GRS 1915+105 curve, $\Gamma(\nu)$ (see Fig. 7) into
$\Gamma(\nu^{\prime})$ produces a fit for the XTE 1550-564 data. This means
that if two sources are in the same accretion regime [i.e. when $\tau_{0}(\gamma)$
is the same] their relative BH masses can measured by the simple shift
transformation $\nu_{low}^{\prime}=m_{1}\nu_{low}/m_{2}$. In our case we use
$m_{1}=12$ for GRS 1915+105 and $m=10$ for XTE 1550-564. These values are very
close to the values that have been obtained by Shrader \& Titarchuk (2003) for
these sources using X-ray spectroscopic methods (Shrader \& Titarchuk 1999).
Thus the TL model provides an independent way to estimate the mass of one or
more BH's in the same accretion state.

A comparison of all data points of  correlating photon index vs QPO low
frequency for XTE 1550-564 (black points) with the inferred correlation is
shown in Figure 11. 
 The saturation index value in the theoretical curve is related to
specific value of the plasma temperature of the converging inflow $kT_{ci}$,
which changes from 5 keV to 20 kev from the top curve to the bottom
respectively. Slightly different cooling regimes of the site of the converging
bulk inflow can explain the different saturation levels of the index observed
in XTE 1550-564. The temperature of the converging inflow is determined by the
disk illumination geometry and by the intrinsic heating of the flow by pairs
produced very close to the BH event horizon (see Laurent \& Titarchuk 2004). All these
effects may be manifested by the observed index-frequency correlation at high
values of $\nu_{low}$.

~~~~~~~~~~~~~~~~~~~~~~~~~~~~~~~~~~~~~~~~~~~~~~~~~~~~~~~~~~~~~~~~~~~~~~~
~~~~~~~~~~~~~~~~~~~~~~~~~~~~~~

\centerline{\it 4U 1630-47}

Figure 12 shows a plot of index versus QPO frequency for 4U 1630-47 [Trudolyubov et al. (1999); Tomsick
\& Kaaret (2000), Kalemci (2002)]. The data again show a similar behavior to
that of XTE J1550-564 with an index saturation at $\Gamma=2.45$ and
$\Gamma=2.65$. One can see two phases with the index level at
$\Gamma=1.7\pm0.1$ and $\Gamma=2.6\pm0.1$ and a transition between them that
is much sharper than that observed in the other sources dicussed above. The best-fit curve
$\Gamma(\nu_{L})$ (solid line) is obtained using $\tau_{0}=1.55\gamma^{0.3}$ for $\gamma\leq \gamma_{\ast}=3.6$ and 
 $\tau_{0}=1.55\gamma_{\ast}^{-0.55}\gamma^{0.85}$ for $\gamma > \gamma_{\ast}$
and $\Gamma(\tau_0)$ function (see formula 14) for the best-fit BH mass $m=16$
which is close to the estimate made by Borozdin et al. (1999) using the X-ray
spectroscopic method. The solid and dash curves are for saturation index value
$\Gamma=2.65$ (related to $kT_{ci}=12$ keV) and $\Gamma=2.45$ (related to $kT_{ci}=20$ keV) respectively.
Coefficients $b_{0}=2.45, ~2.65$, $b_{1}=53.7, ~44.3$ and $\tau_{\star}=0.55, ~0.65$ 
for $\tau_{0}>2.2 $ and $\tau_{0}>3.42 $  for solid and dash curves respectively. 

 Our explanation for the observed anomalously sharp transition of the spectral index is as follows.
Because at the sharp spectral transition QPO frequencies do not change  
we can speculate that  the TL size also  does not change at the sharp transition. But if the TL size 
is almost the same then the $\gamma-$parameter is almost the same because the TL size depends on $\gamma-$
parameter only (see section 2.1). But the spectral index is a function of $\tau_0$ and the TL plasma
temperature. If the index rises  drastically this means that upscattering is
also drastically suppressed  during the transition.  Then we have only one possibility i.e. that we are observing 
a sharp switching from the thermal Comptonization to the Bulk Motion Comptonization regimes. 
At the end of this transition we  see a signature of the Bulk inflow in terms of  the index  saturation. 
The Bulk Motion Comptonization is
dominant when the plasma temperature drops significantly to the value of order 10 keV. This  can  happen 
when a strong cooling emerges  either because  the  optical depth $\tau_0$ increases due the accumulation
matter in the transition layer or because an additional source of cooling such as a soft photon supply from the disk
appears (see investigation of these cases in section 2.2). The latter case can be realized if the corona intercepts
more disk soft photons due  to a puffing up of the disk  or a build up of the disk beneath the corona.

\section{Discussion and Summary of the Results}

We have developed a model of accretion onto a Black Hole which greatly
simplifies and reclassifies the plethora of ``states'' observational assigned
to categorize the X-ray observations of variable BHC into two generic phases
(states):

I) \textit{soft state} where we see the effect of the BH as a
\textquotedblleft drain\textquotedblright. Bulk inflow upscattering  of disk photons dominates the behavior
of the BH spectrum. The power law spectrum is steep in this situation. The
observed high energy photons are emitted from a compact region, where soft
energy photons of the disk are upscattered by bulk matter inflow forming the
steep power law with photon index around 2.8, and low QPO frequencies (above
10 Hz) and high QPO frequencies (of the order of 100 Hz) are observed. In
terms of the relativistic particle acceleration in the region of the corona
and its effect on upscattering the soft disk photons, the soft state spectrum
is a result of the first order Fermi acceleration with respect to $V/c$, i.e. 
 a relative photon energy change $\Delta E/E\propto V/c$.
We would like to emphasize that bulk inflow is present in BH when the high
mass accretion is high \textit{but not in NS}, where the presence of the firm
surface leads to the high radiation pressure which eventually stops the
accretion. The bulk inflow and all its spectral features are absent in NSs, in
particular, the saturation of the index $\Gamma\sim2.8$ with respect to QPO
frequency, which is directly related to the optical depth and mass accretion
rate, observed in the soft spectral state of only BHs and is therefore a
particular signature of a BH.

II) \textit{hard state,} which is comparatively starved for accretion. The
hard phase (state) is related to an extended thermal Compton scattering cloud
(cavity) characterized by a photon index $\Gamma\sim 1.7$ and the
presence of low QPO frequencies (below 1 Hz). The low-hard spectrum is a result
of the Fermi acceleration of the second order with respect to $V/c$, i.e.
$\Delta E/E\propto (V/c)^2$. The
effect of the first order on $V/c$ is smeared out by the quasi-symmetry of the
particular dynamic, predominantly thermal motion of the Compton cloud plasma.

The Transition Layer Componization Model: i. specifies the precise behavior of
the power law index for each state; ii. identifies the origin and nature of low
frequency and high frequency QPOs; iii. predicts and explains the observed
relationship between them and their dependence mass accretion rate and
spectral photon index for each of the phases; iv. predicts and explains the
observed robust correlation of low frequency QPOs with spectral index observed
in BHCs; v. predicts and explains the observed correlation between low and
high frequency QPOs in BHs, neutron stars and white dwarfs. 
\ We interpret the
correlation between low frequency $\nu_{low}$ and power-law photon index
$\Gamma$  investigated by Sobczak et al. (1999), (2000); 
Remillard (2000b); Homan et al. (2001); Kalemci (2002),  Vignarca et al. (2003)
 and Fiorito, Markwardt \& Swank (2003)   for a variety of BH sources and states. 
The observed
correlation strongly constrains theoretical models and provides clues to
understanding the nature of the QPO phenomena particularly in BHs and in
compact objects in general. 
  
The key assumptions and limitations of the model are: 

i. a sub-Keplerian rotation of central object (NS and BH) 
which is observationally well established fact in NSs of LMXB but it still needs  verification for BHs.
However the existence of an adjustment region i.e. the TL is quite reasonable assumption for all but very rapidly rotating
BHs. Unfortunately up to now there is no direct measurement of a BH spin.  

ii. Our fit to the data uses  the model predicted  dependence of the index on the TL optical depth $\tau_0$
and the model predicted dependence of the QPO frequencies on the $\gamma-$parameter. It is clear from the model that $\tau_0$
and $\gamma$ are correlated but within TLM model one can not determine the exact $\tau_0(\gamma)-$dependence
because of the uncertainty of the disk equation of the state and the viscosity in the disk. 
Therefore we have determined  the $\tau_0(\gamma)$ relation by fitting  the observed data, i.e. the  correlation
between the QPO low-frequency and the photon spectral index.  
As a result of our analysis of the observed index-QPO correlation we conclude  that the  function
$\tau_0(\gamma)$ is almost the same  for GRS 1915+105 and XTE J1550-564 but it is  drastically different for 4U 1630-47.
For this source the inferred $\tau_0$ shows a very strong dependence of $\gamma$ on the QPO  frequencies produced 
during a
sharp spectral state transition. We speculate that this peculiar behavior of the index vs QPO frequency and consequently
the inferred $\tau_0$ vs $\gamma$ is a result of the plasma accumulation in the transition layer. It is possible that a
similar index-frequency correlation  can be detected in other sources.   

iii. The model fits to the data gives us the relatively low values of $\gamma=(2-20)$ (corresponding to  viscosity
parameter $\alpha=\gamma^{-1}$ from to 0.05 to 0.5). Such low values of $\gamma$ can be understood in the framework of MHD treatment
of the disk viscosity. In fact, Soria (1999) inferred  $\alpha$ of order 0.1 (regardless of the true disk viscosity) 
and higher if in the disk  a magnetic pressure is 10\% of the gas pressure or higher. 

iv. Within TLM  one can estimate the absolute normalization of the magnetoacoustic (QPO low) frequencies (see e.g. TBW01) 
but a more precise normalization can be only obtained  using the observed low-high frequency correlation (see Fig. 2).

v. In some very soft (or extended power law) regimes the
positive correlation of frequency and index  appears to be
unbounded, i.e. there is no indication of saturation of the index value (FMS).
Explanation for this effect is out of  scope of the present TL model. We can only suggest here that 
 the extended power law regime occurs when  strong outflows obscure the bulk inflow
and the relatively cold outflow from winds downscatters the emerging high energy photons and soften the observed spectrum 
(we will provide the details of this picture  elsewhere). 

vi. Vignarca's Figure 6 plots the index$-$frequency for two sources XTE J1550-564
and GRO J1655-40 together. The behavior of the later sources is quite distinct
from all others analyzed - it shows a reverse or \textit{negative} correlation
between photon index and frequency. This is the only case known to exhibit
this type of behavior. We  propose that at a high accretion rate regime 
 the  pair production heating adds to and can dominate the heating in the cavity causing a transition back
to thermal Comptonization phase. In this phase a reverse correlation between
low QPO frequency can result and is our explanation of the single case where
this phenomena is observed. Further theoretical investigation of this case will be addressed in future studies.

  The main results of this paper are: 

(1) We find that the observed low frequency - the index correlation is a 
natural consequence of an adjustment of the Keplerian disk flow to the
innermost sub-Keplerian boundary conditions near the central object. This
ultimately leads to the formation of the sub-Keplerian transition layer (TL)
between the adjustment radius and the innermost boundary (the horizon for BH).

(2) In the framework of the TL model $\nu_{high}$ is related to the
gravitational frequency at the outer (adjustment) radius $\nu_{g}\approx\nu_{\rm K}$ and
$\nu_{low}$ is related to the magnetoacoustic oscillation frequency $\nu_{MA}
$. Using a relation between $\nu_{MA}$ and the mass accretion rate and the
photon index $\Gamma$ and the mass accretion rate we infer a correlation
$\ $between $\nu_{MA}$ and the spectral index $\Gamma$.

(3) Identification of $\nu_{low}$ with $\nu_{MA}$, allow us to make a
comparison of the theoretically predicted correlation with the observed
correlation. For this identification we use the one temperature plasma
assumption. We apply the plasma temperature obtained from the Comptonization
spectra (electron temperatures) for calculations of magnetoacoustic
frequencies which strongly depend on the proton temperature. The one
temperature assumption is quite consistent with the data.

(4) We present strong arguments that in BHs the two particular distinct phases
occur in which one of them, the steep power-law phase is the signature of a  BH.

(5) We found that a hard phase (state) related to an extended Compton cloud
(cavity) characterized by the photon index around 1.7 and the low QPO
frequencies below 1 Hz. This is the regime where thermal Comptonization
dominates the upscattering of soft disk photons and the spectral shape (index)
is almost independent of mass accretion rate.

(6) We find that the soft phase (state) is related to the very compact region
where soft energy photons of the disk are up scattered forming the steep power
law spectrum with photon index saturating around 2.8. This is the regime where
Bulk Motion Comptonization dominates and the effect of an increase in the mass
accretion is offset by the effect of photon trapping in the converging flow
into the BH.
  
 (7) In the context of distinquishing between NS and BH sources we would like to note  the following  
for low hard states. The source can only be a NS if a soft blackbody-like component with the color temperature 
of order of  1 keV is observed in the spectrum (Torrejon et al. 2004). 
On the contrary, such a high temperature is necessarily related to a high disk luminosity in the case 
of a BH and this case is never observed for BH's 
in the low hard state

(8) We offer a new method of BH mass estimation using the index-frequency
correlations; namely, if the theoretical curve of the index-frequency
dependence $\Gamma(\nu_{low})$ related to the BH mass parameter $m_{1}$ fits the data
for a given source then the simple slide of the frequency axis 
$\nu_{low}^{\prime}=(m_{1}/m_2)\nu_{low}$ with respect to $\nu_{low}$ may allow us to
obtain the mass $m_{2}$ by fit of $\Gamma(\nu_{low}^{\prime})$ to the observed
correlation for another source.


We acknowledge Tomaso Belloni for kindly supplying us data published in Vignarca et al. (2003). 
We also  acknowledge and thank the referee 
for his/her suggestions  to clarify our results and explain the limitations of our model.


\newpage

\begin{figure}[ptb]
\includegraphics[width=6.5in,height=5.in,angle=0]{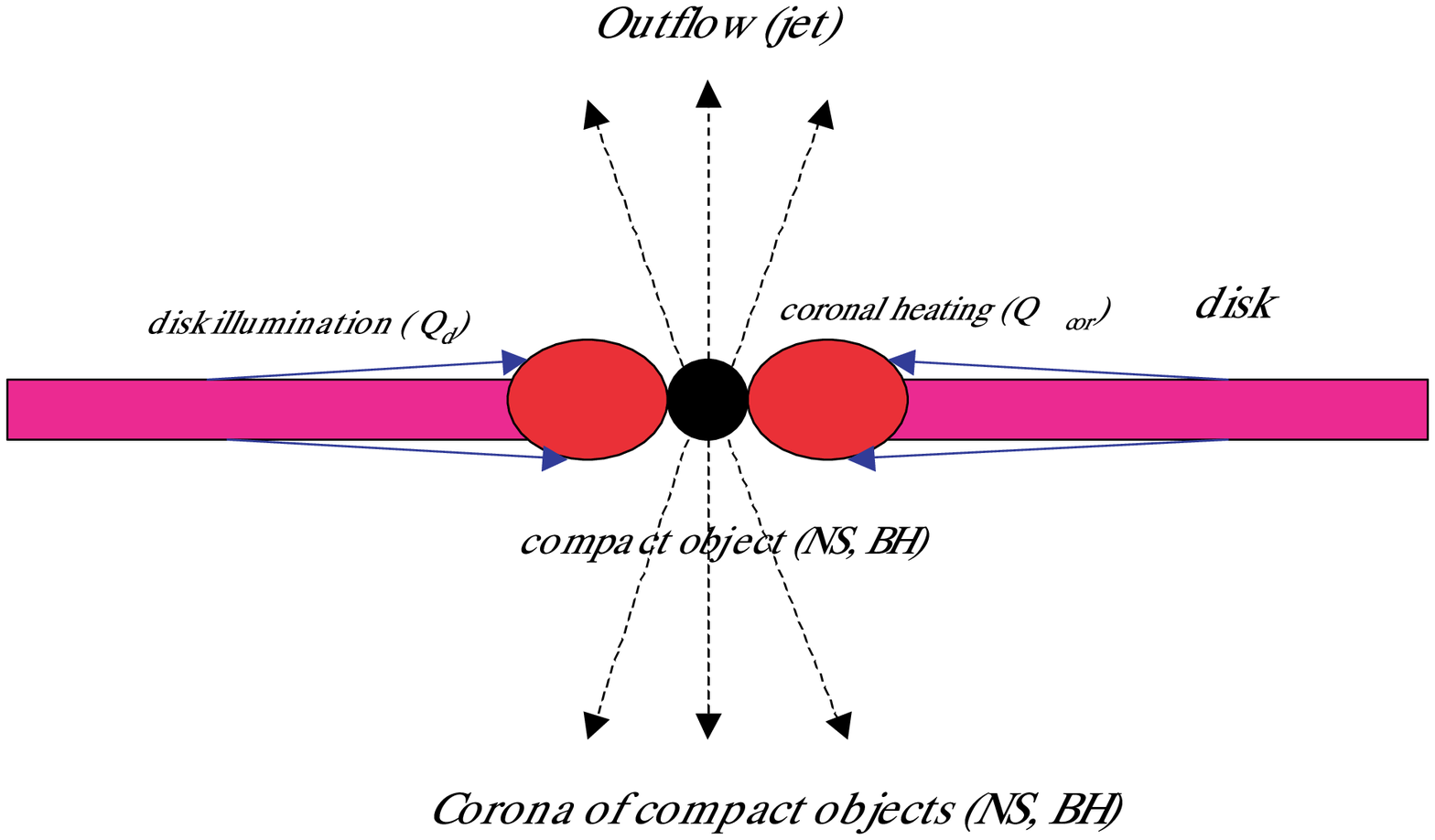}
\caption{ Transition layer (corona) concept. This picture renders the
gravitional energy release in the disk and corona and the coronal illumination
by the disk soft flux in compact object (NS and BH). See details in text }
\end{figure}\clearpage

\begin{figure}[ptb]
\includegraphics[width=5in,height=6in,angle=0]{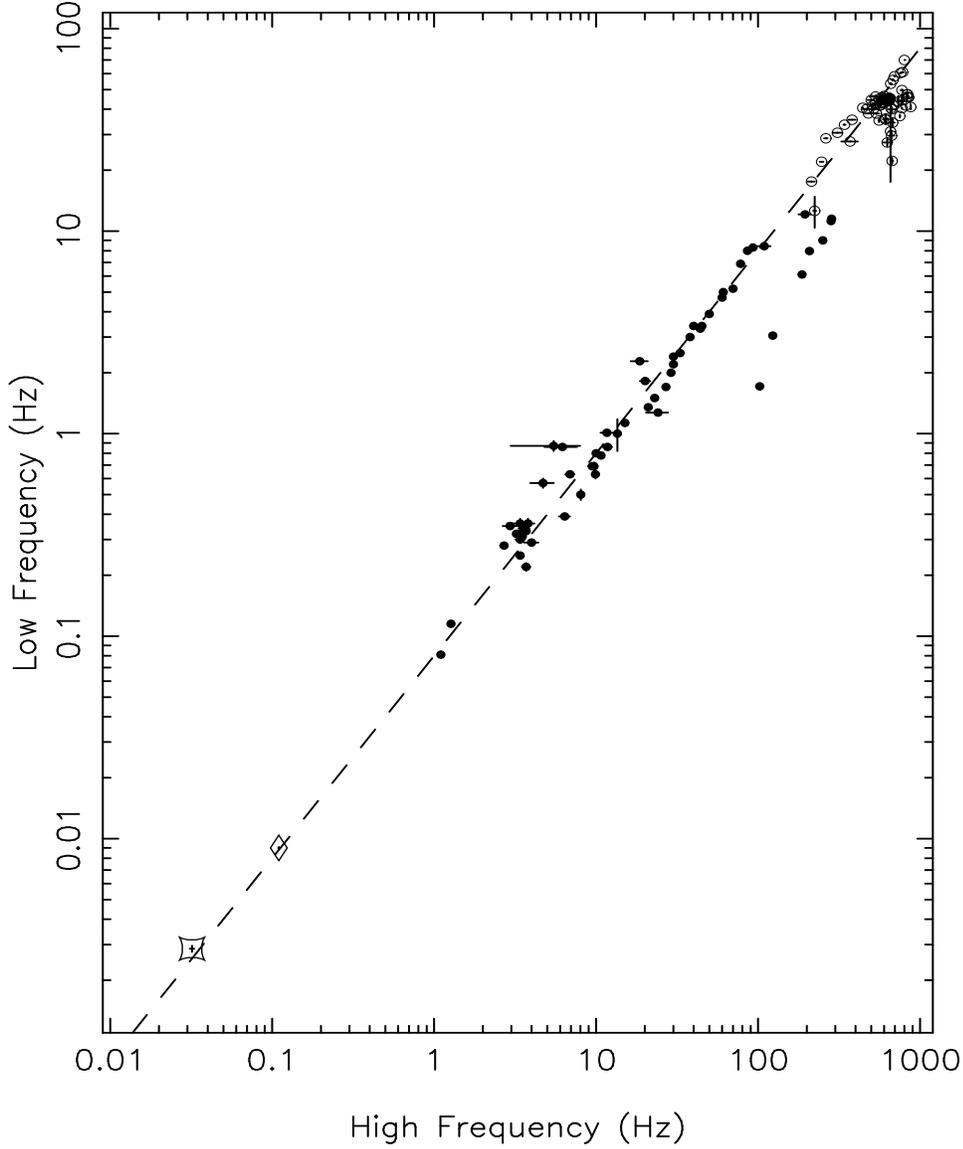}
\caption{Correlation between frequencies of QPO and noise components white
dwarfs [diamonds for SS Cyg (Mauche 2002); squares for VW Hyi (Woudt \& Warner
2002)] , neutron star (open circles) and, black hole candidate (filled
circles) sources. Neutron star and black hole data are from Belloni, Psaltis
\& van der Klis (2002). The dashed line represents the best-fit of the
observed correlation (see TBW01 and TW02 where this correlation is predicted and explained using the TL model).  
This plot also appears in Mauche (2002).}
\label{fig10}
\end{figure}

\begin{figure}[ptb]
\includegraphics[width=5in,height=6in,angle=-90]{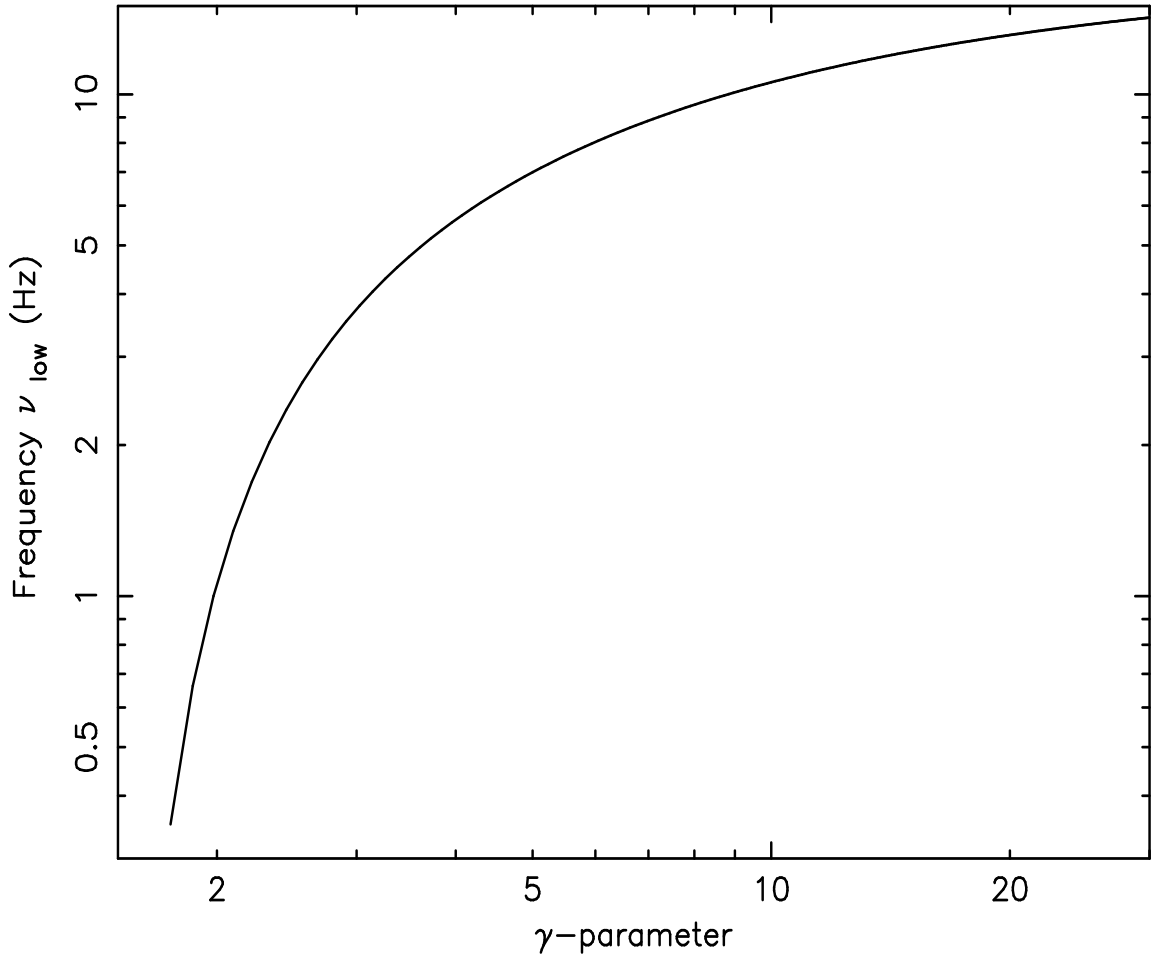}
\caption{Plot of
QPO low frequency vs. $\gamma$ parameter ($\propto\dot M$). }
\label{nug}%
\end{figure}

\begin{figure}[ptb]
\includegraphics[width=5in,height=6in,angle=0]{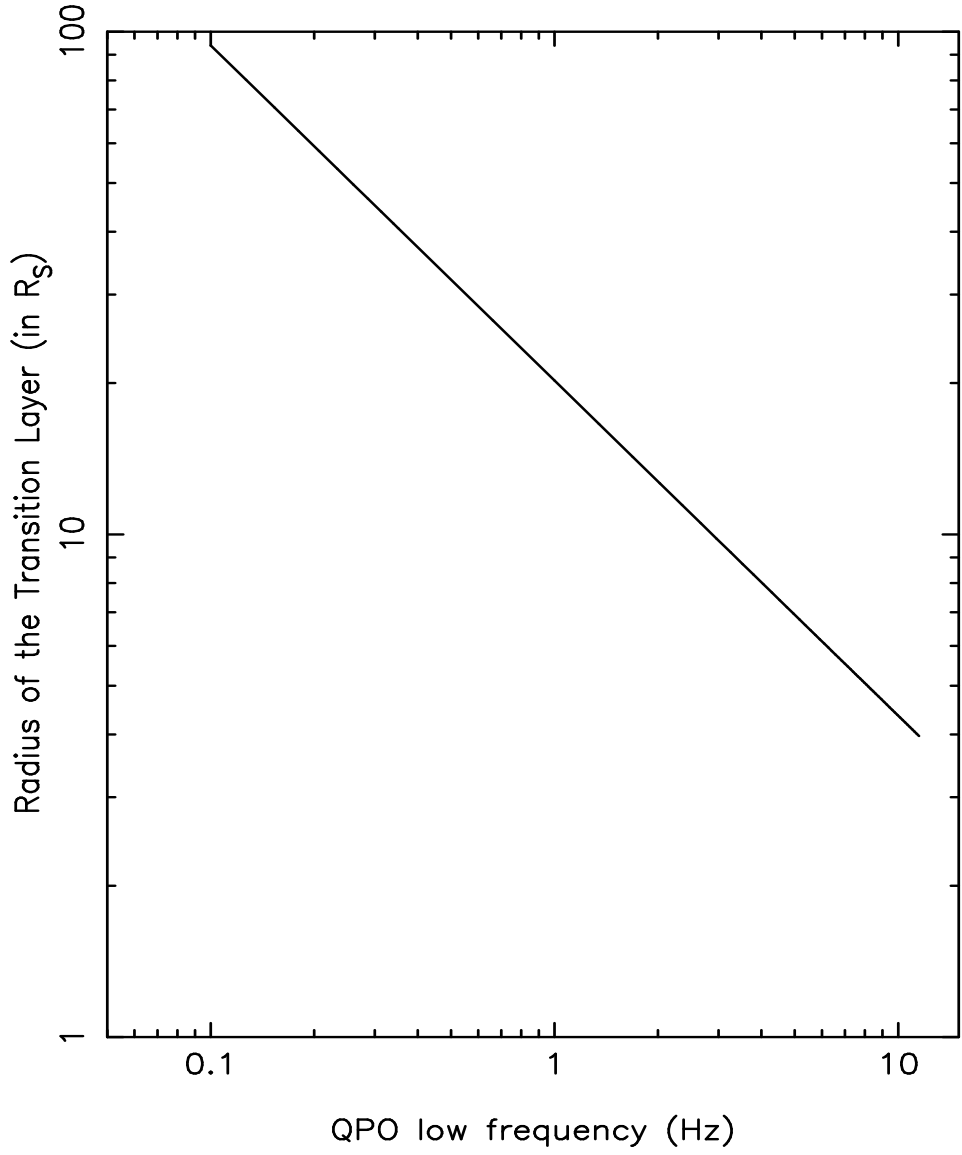}\caption{The outer (adjustment) radius of
the transition layer vs QPO low frequency. }
\label{rnu}
\end{figure}\begin{figure}[ptbptb]

\includegraphics[width=5in,height=6in,angle=0]{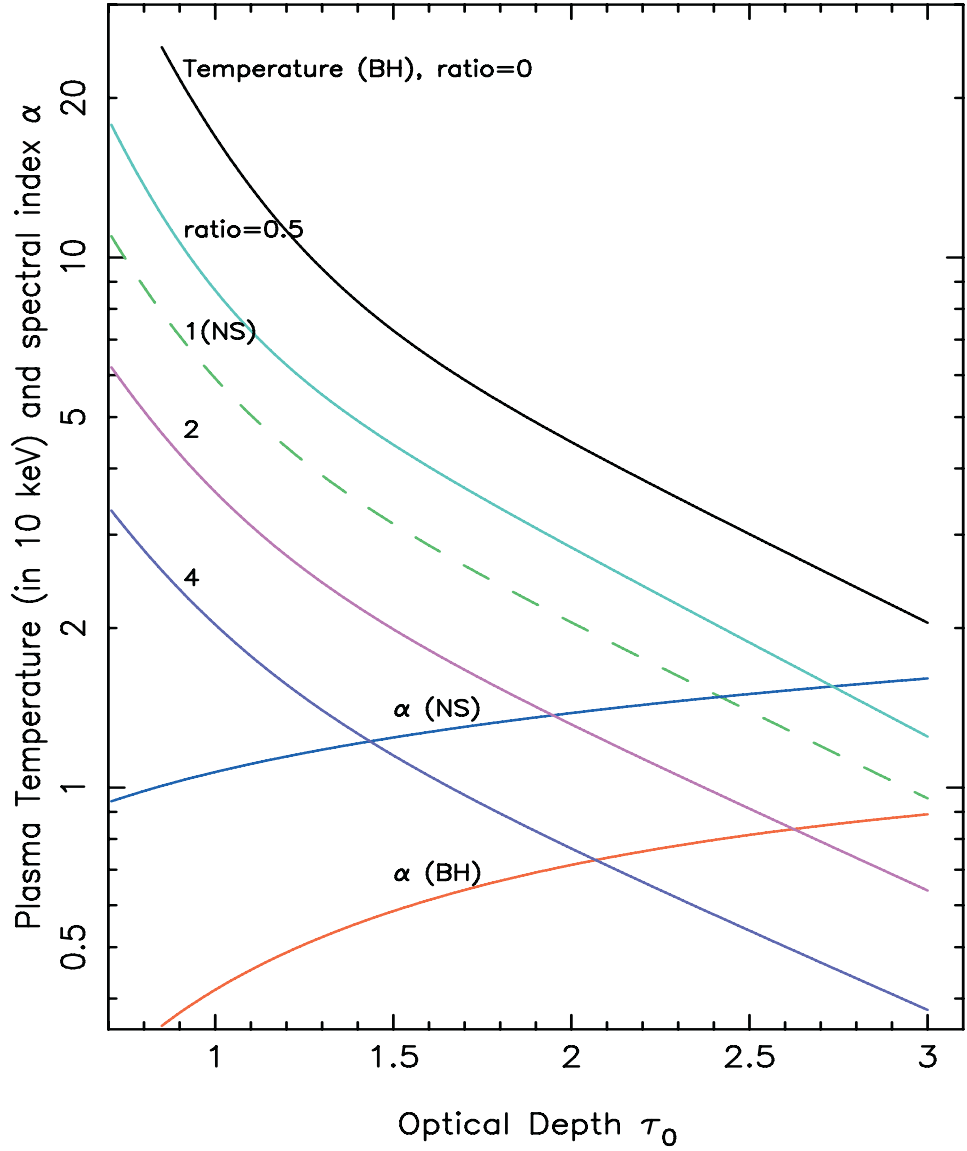}\caption{Energy
spectral index vs the TL optical depth. (Thermal Comptonization case) }
\label{indtauc}
\end{figure}

\begin{figure}[ptbptbptb]
\includegraphics[width=5in,height=6in,angle=-90]{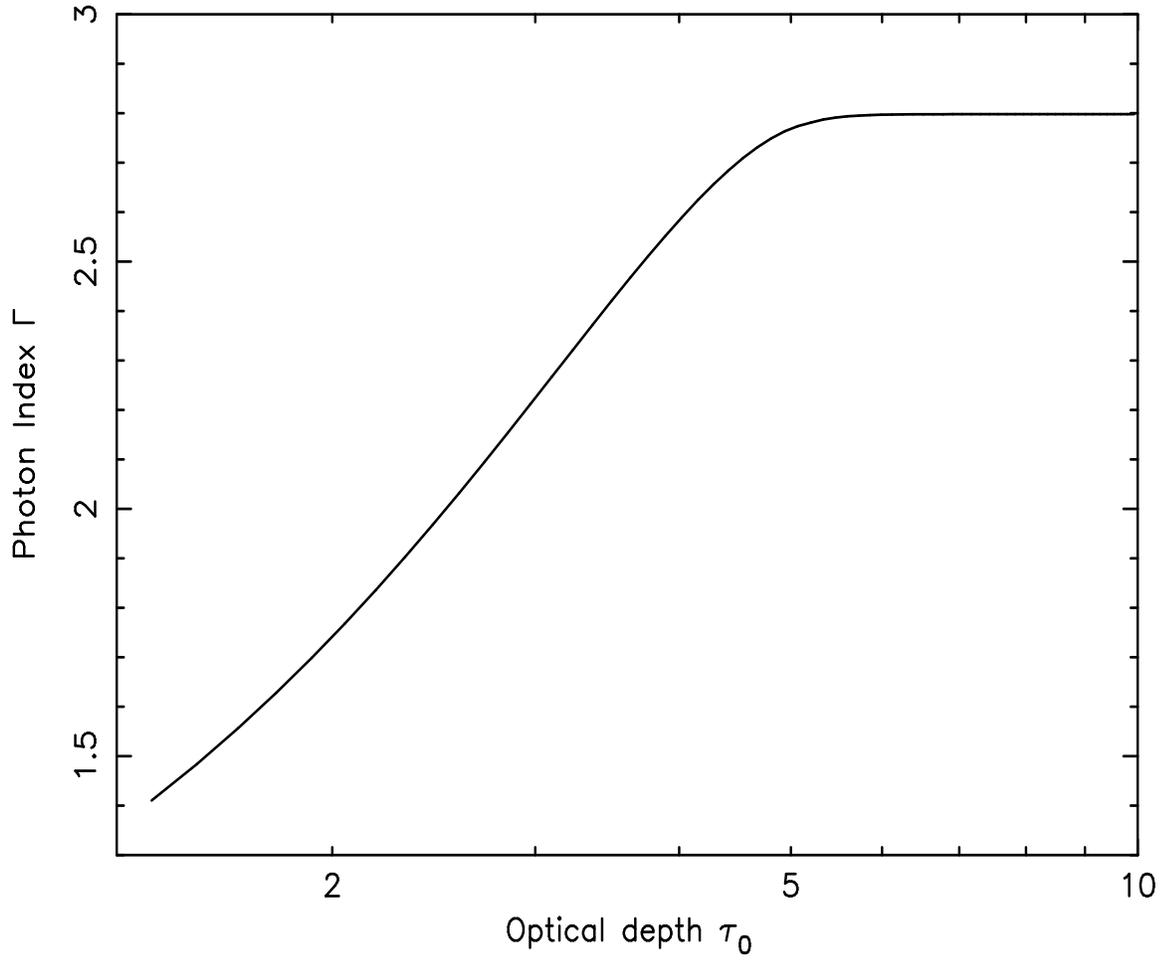}
\caption{Photon
spectral index vs the TL optical depth. (General case) }
\label{indtaug}
\end{figure}


\begin{figure}[ptb]
\includegraphics[width=5in,height=6in,angle=-90]{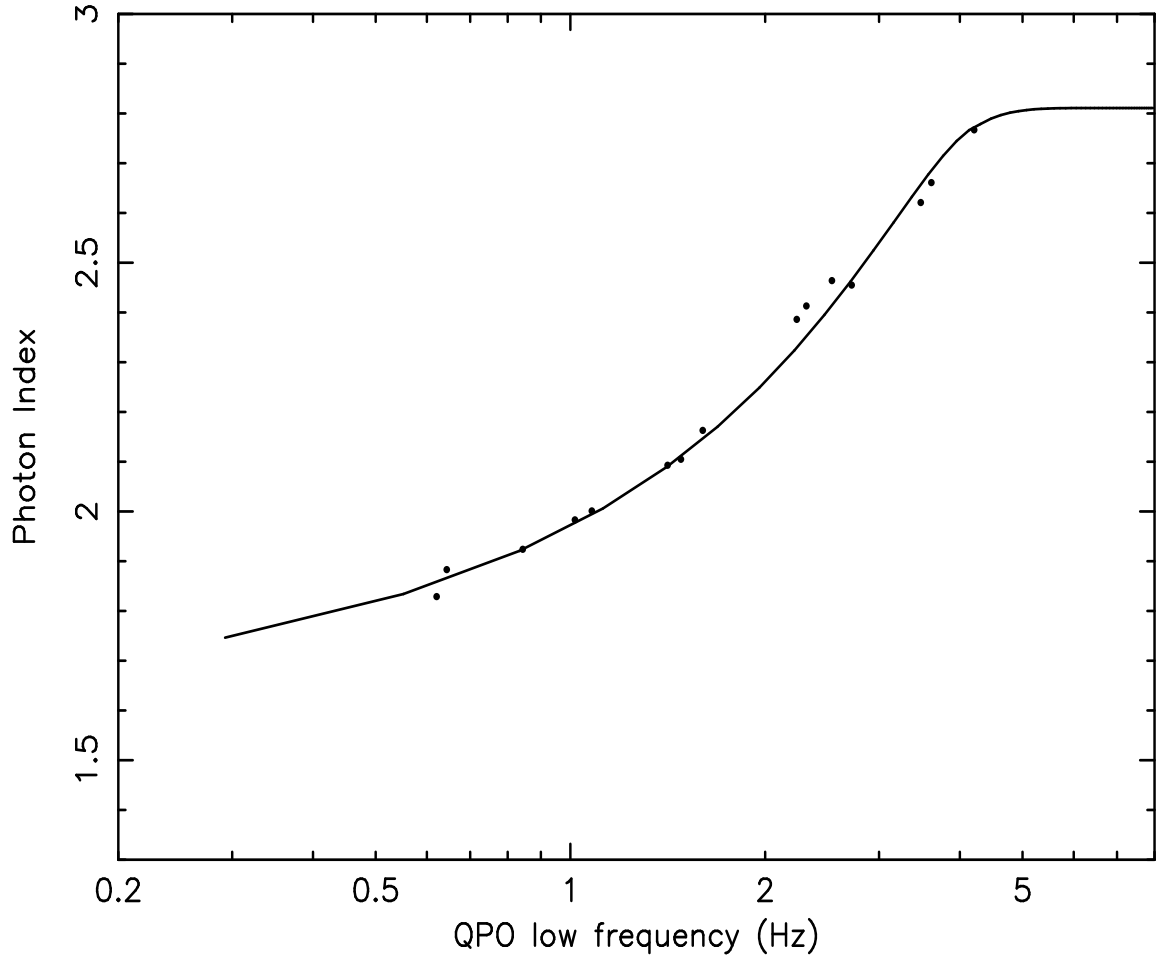}
\caption{ Plot of power-law index versus QPO centroid frequency for the plateau observations
of GRS 1915+105 from Vignarca et al. (2003) along with a fit using the TL model
with $m=12$ and $\tau_{0}=\gamma^{1.25}$. }
\label{plat}
\end{figure}

\begin{figure}[ptb]
\includegraphics[width=6in,height=6in,angle=0]{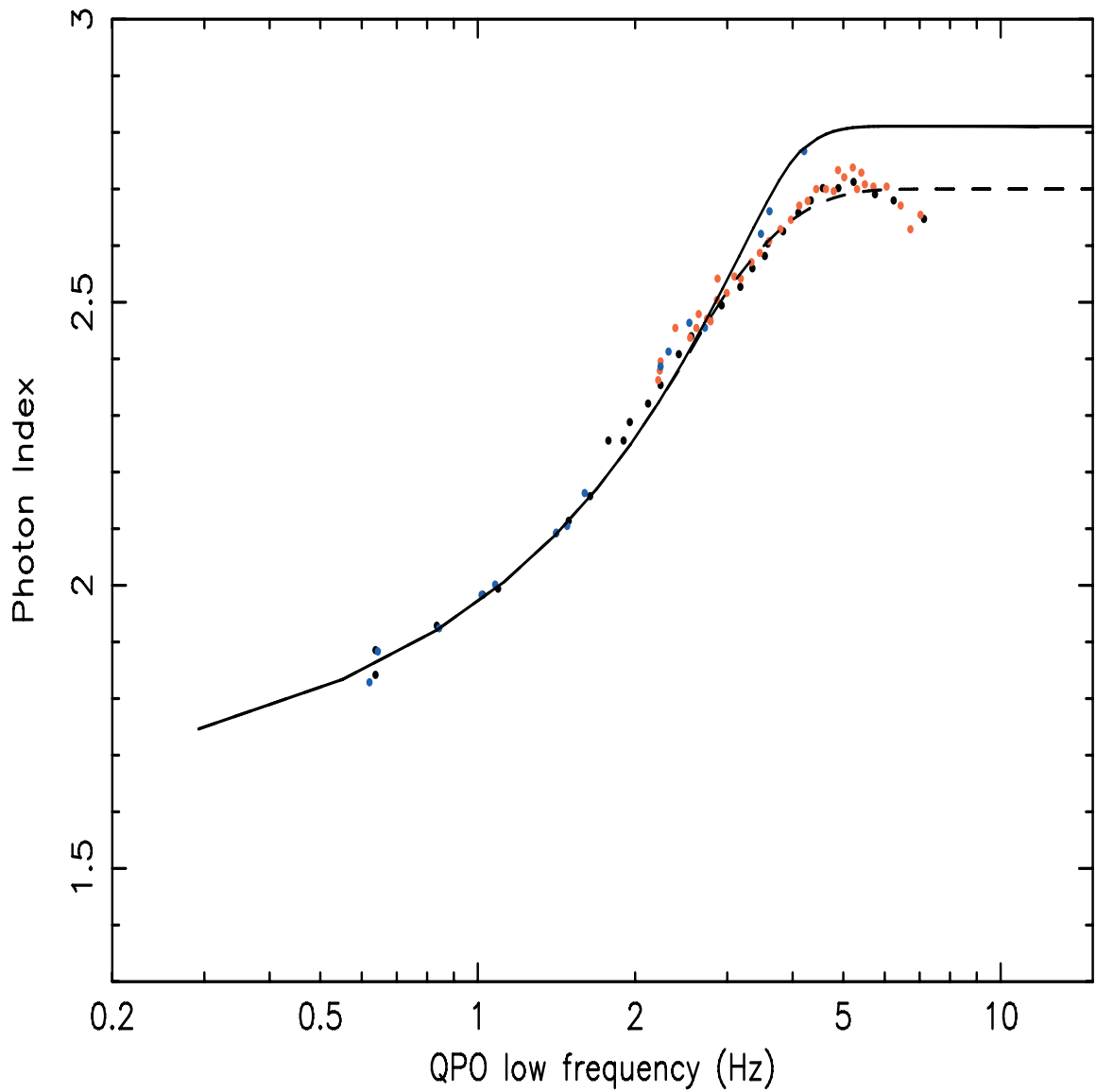}
\caption{ Plot of power-law index versus QPO centroid frequency for the observations of class
$\beta$ and $\nu$ (black points) and $\alpha$ and $\nu$ (red points =obs.
15,16) of GRS 1915+105 from Vignarca et al. (2003) along with a fit using the TL
model with $m=12$ and $\tau_{0}=\gamma^{1.25}$. Values for the plauteau
observations (see previous Figure) are plotted for comparison
(blue points). }
\label{top}
\end{figure}

\begin{figure}[ptbptb]
\includegraphics[width=6in,height=6in,angle=0]{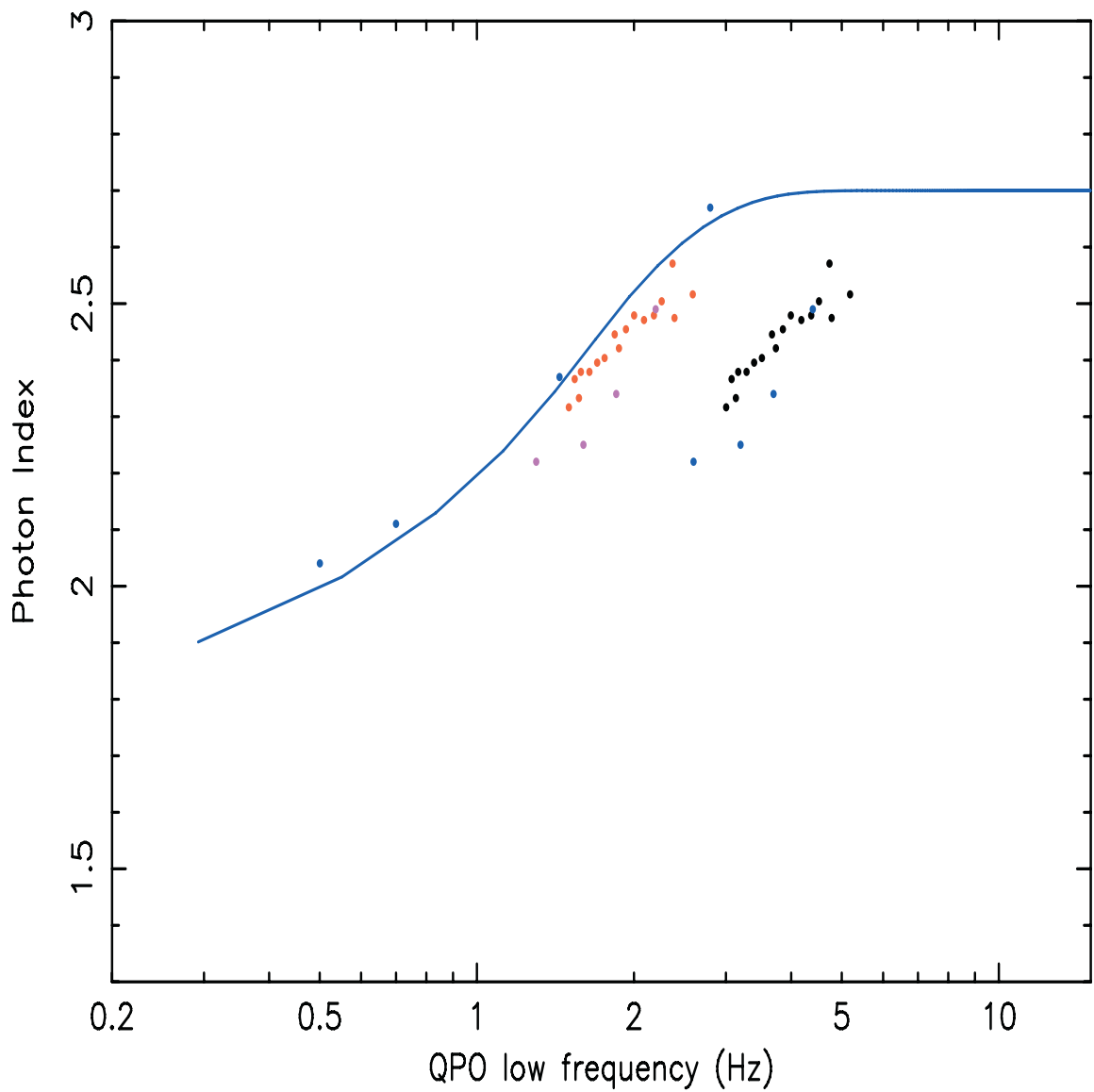}
\caption{ Plot
of power-law index versus QPO centroid frequency for the observations of class
$\alpha$ and $\nu$ of GRS 1915+105 from Vignarca et al. (2003) (black
points=obs. 18). Blue points correspond to the values for observations by
Fiorito et al (2003) (see text). Magenta points correspond to positions where
should be points with half of those frequencies. Red points correspond to
positions where should be points with half of frequencies for obs. 18 (red
points). A curve (blue solid line) is for $m=12$ and $\tau_{0}=\gamma^{1.5}$.
}
\label{bot}
\end{figure}

\begin{figure}[ptbptbptb]
\includegraphics[width=6in,height=6in,angle=0]{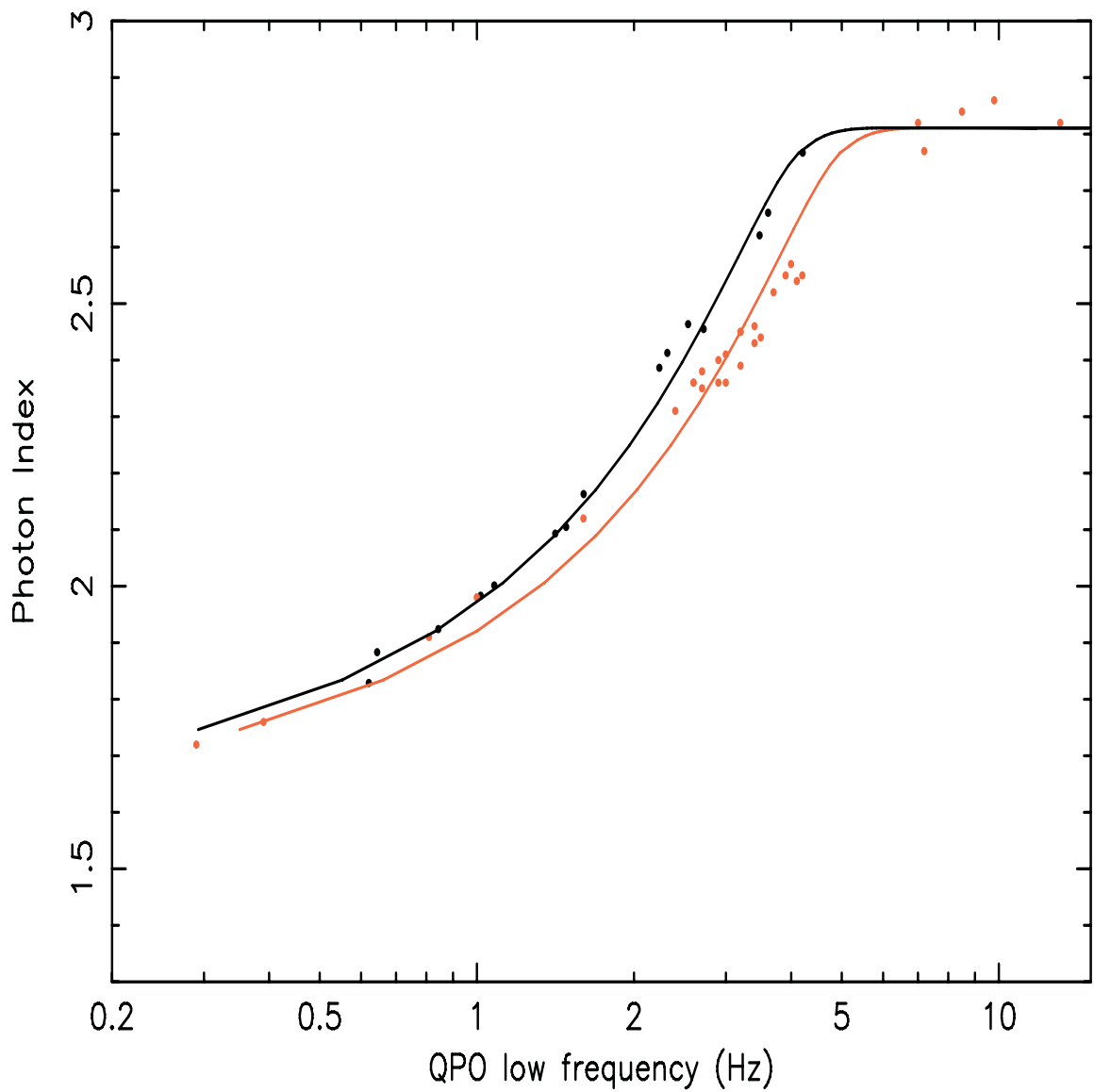}
\caption{
Comparison of the observed (points) and the  theoretical correlations
(solid lines) of photon index vs QPO low frequency between GRS 1915+105 (observations by Vignarca et al. 2003) and
XTE J1550-564 [observations by Sobczak et al. (1999), (2000); Remillard et al. (2002a,b), see also 
  Fig. 6, 8 in Vignarca et al.]. Black points and line for GRS 1915+105 and red points and line
for XTE J1550-564. The XTE J1550-564 curve is produced by sliding the GRS
1915+105 curve along the frequency axis with factor $12/10$ (see text for
details). }
\label{1915_1550}
\end{figure}

\begin{figure}[ptb]
\includegraphics[width=6in,height=6in,angle=0]{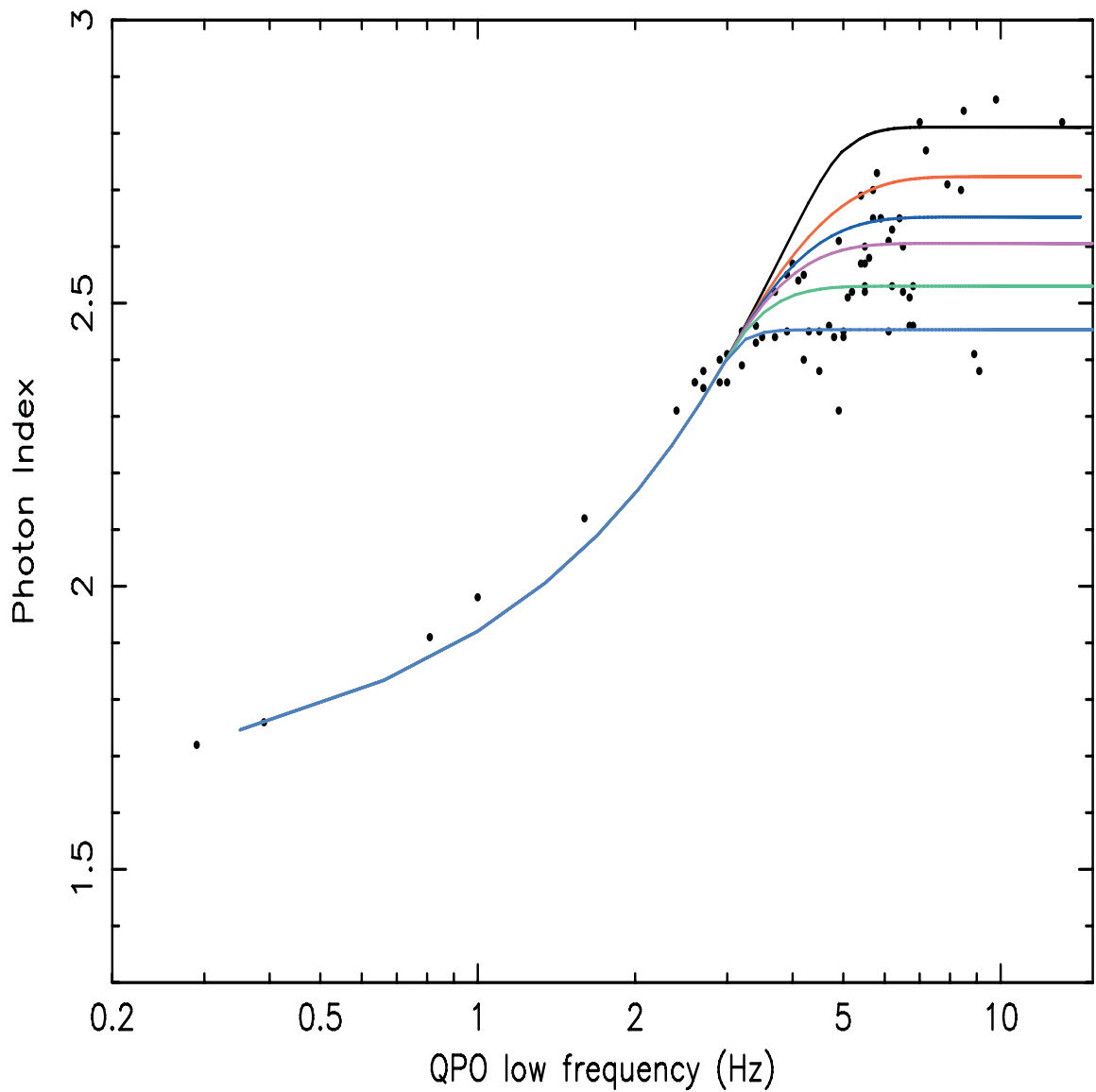}\caption{
Comparison of the observed correlation of photon index vs QPO low frequency
for XTE 1550-564  by Sobczak et al. (1999), (2000); Remillard et al. (2002a,b),  (black points) 
with the inferred correlation The saturation
index value in the theoretical curve is related to specific value of the
plasma temperature of the converging inflow $kT_{ci}$, which changes from 5
keV to 20 kev from the top curve to the bottom respectively. }
\label{1550}
\end{figure}

\begin{figure}[ptb]
\includegraphics[width=6in,height=6in,angle=0]{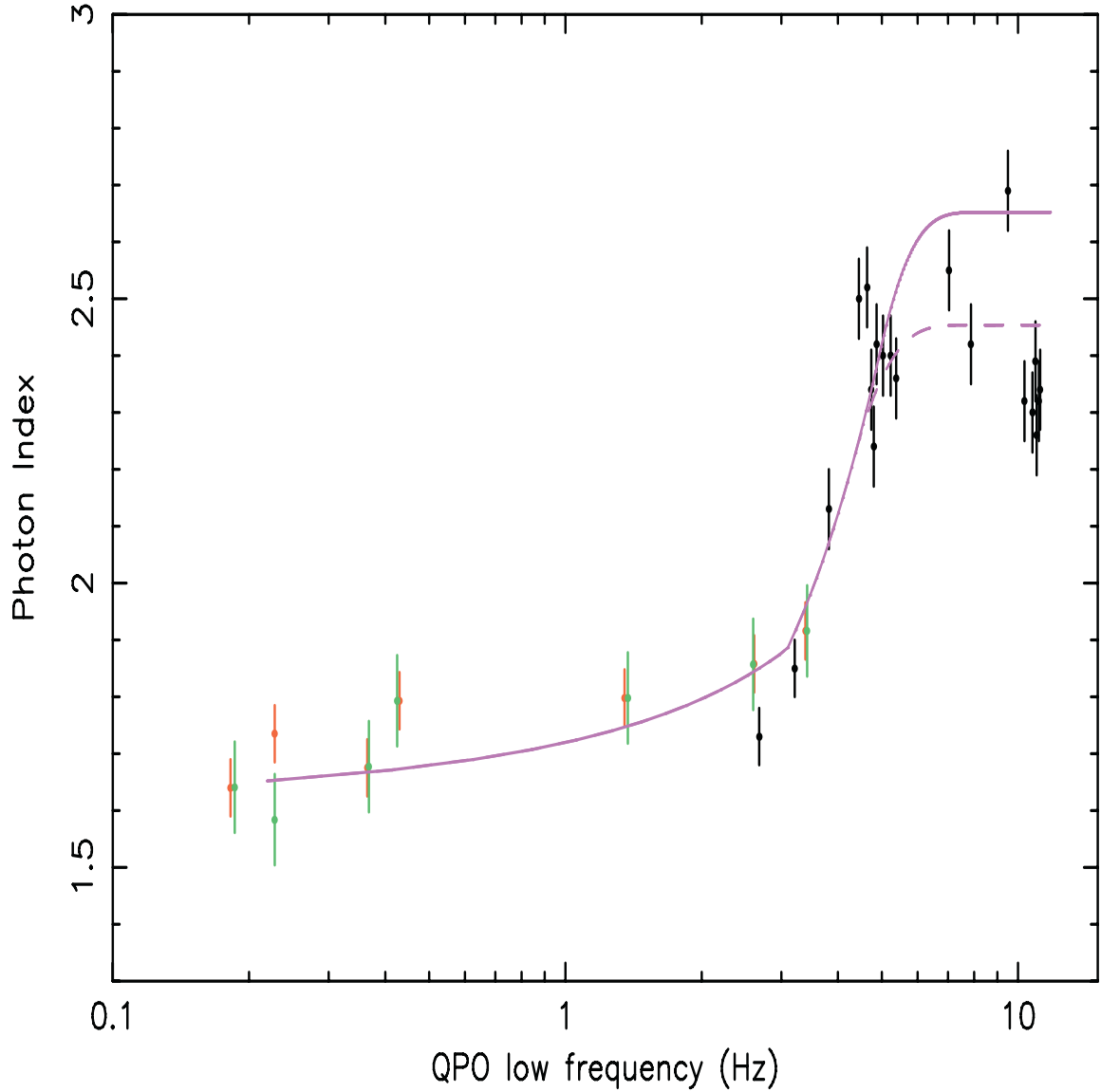}
\caption{
Comparison of the observed correlation of photon index vs QPO low frequency or 4U 1630-47 [Trudolyubov et al. (1999); 
Tomsick \& Kaaret (2000) and Kalemci (2002)]  with the inferred correlation. The saturation
index value in the theoretical curve is related to specific value of the
plasma temperature of the converging inflow $kT_{ci}$, which changes from 12
keV to 20 kev from the top to the bottom  respectively.}
\end{figure}

\end{document}